\documentclass[numberedappendix,iop,twocolappendix,superscriptaddress]{emulateapj} % May yield bad PDF, good DVI
\usepackage{epsfig}
\usepackage{verbatim} % needed for comments
\usepackage{graphicx}
\usepackage{tablefootnote}
\usepackage{natbib}
\usepackage{dcolumn}% Align table columns on decimal point
\usepackage{bm}% bold math
\usepackage{epsf}
\usepackage{amsmath}
\usepackage{amssymb}
\usepackage{amsfonts}
\usepackage{appendix}
\usepackage{xcolor}
\usepackage{soul}
\usepackage[normalem]{ulem} % needed for strikeout font
\usepackage{pifont}
\usepackage{multirow}
\usepackage[percent]{overpic}

\definecolor{darkgreen}{rgb}{0.0,0.5,0.0}
\usepackage[colorlinks,citecolor=darkgreen]{hyperref} % Works with PDFLaTeX; use for arXiv and ShareLaTeX.

\newcommand{\ie}{\emph{i.e.,} }
\newcommand{\eg}{\emph{e.g.,} }

\newcommand{\be}{\begin{equation}}
\newcommand{\ee}{\end{equation}}
\newcommand{\bea}{\begin{equation*}}
\newcommand{\eea}{\end{equation*}}
\newcommand{\beqr}{\begin{eqnarray} \nonumber}
\newcommand{\eeqr}{\end{eqnarray}}
\newcommand{\beqrb}{\begin{eqnarray}}
\newcommand{\eeqrb}{\nonumber \end{eqnarray}}
\newcommand{\fin}{\mbox{ .}}
\newcommand{\coma}{\mbox{ ,}}

\newcommand{\cm}{\mbox{ cm}}
\newcommand{\sr}{\mbox{ sr}}
\newcommand{\se}{\mbox{ s}}
\newcommand{\yr}{\mbox{ yr}}

\newcommand{\Gyr}{\mbox{ Gyr}}
\newcommand{\erg}{\mbox{ erg}}

\newcommand{\km}{\mbox{ km}}

\newcommand{\Mpc}{\mbox{ Mpc}}

\newcommand{\keV}{\mbox{ keV}}
\newcommand{\MeV}{\mbox{ MeV}}
\newcommand{\GeV}{\mbox{ GeV}}
\newcommand{\TeV}{\mbox{ TeV}}

\newcommand{\const}{\mbox{const.}}

\newcommand{\gama}{$\gamma$}

\newcommand{\dgr}{{^\circ}}
\newcommand{\till}{{\mbox{--}}}

\newcommand{\mass}{\bar{m}}

\newcommand{\mdot}{{\dot{m}}}

\newcommand{\eps}{{\epsilon}}

\newcommand{\myTS}{\mbox{TS}}

\newcommand{\varrhoNorm}{{\tilde{\varrho}}}

\newcommand{\model}{\mu}
\newcommand{\Mbin}{\mathcal{M}}

\def\myfig#1{#1}

\defcitealias{KeshetEtAl12_Coma}{K17}
\newcommand{\Coma}{{\citetalias{KeshetEtAl12_Coma}}}

\defcitealias{ReissEtAl17}{R17}
\newcommand{\Stack}{{\citetalias{ReissEtAl17}}}

\begin{document}

\title{X-ray to gamma-ray virial shock signal from the Coma cluster}

%\correspondingauthor{Uri Keshet}
%\email{ukeshet@bgu.ac.il}
%\author{Uri Keshet}
%\affiliation{Physics Department, Ben-Gurion University of the Negev, Be'er-Sheva 84105, Israel}
%\author{Ido Reiss}
%\affiliation{Physics Department, Ben-Gurion University of the Negev, Be'er-Sheva 84105, Israel}
%\affiliation{Physics Department, Nuclear Research Center Negev, POB 9001, Be’er-Sheva 84190, Israel}

\author{
Uri Keshet\altaffilmark{1}
and
Ido Reiss\altaffilmark{1,2}
}

\email{ukeshet@bgu.ac.il}

\altaffiltext{1}{
Physics Department, Ben-Gurion University of the Negev, Be'er-Sheva 84105, Israel
}

\altaffiltext{2}{
Physics Department, Nuclear Research Center Negev, POB 9001, Be’er-Sheva 84190, Israel
}

\shorttitle{Virial X-rays to \gama-rays in Coma}
\shortauthors{Keshet \& Reiss}

\begin{abstract}
Following evidence for an east--west elongated virial ring around the Coma galaxy cluster in a $\sim220$ GeV VERITAS mosaic, we search for corresponding signatures in $>$GeV $\gamma$-rays from Fermi-LAT, and in soft, $\sim0.1$ keV X-rays from ROSAT.
For the ring elongation and orientation inferred from VERITAS, we find a $3.4\sigma$ LAT excess, and detect ($>5\sigma$) the expected signature in ROSAT bands R1 and R1+R2.
The significances of both LAT and ROSAT signals are maximal near the VERITAS ring parameters.
The intensities of the ROSAT, Fermi, and VERITAS signals are consistent with the virial shock depositing $\sim0.3\%$ (with an uncertainty factor of $\sim3$) of its energy over a Hubble time in a nearly flat, $p\equiv - d\ln N_e/d\ln E\simeq 2.0$--$2.2$ spectrum of cosmic-ray electrons.
The sharp radial profiles of the LAT and ROSAT signals suggest preferential accretion in the plane of the sky, as indicated by the distribution of neighboring large-scale structure.
The X-ray signal gauges the compression of cosmic-rays as they are advected deeper into the cluster.
\end{abstract}

%\maketitle

%\sloppy
\keywords{galaxies: clusters: individual (Coma) --- gamma rays: galaxies: clusters --- X-rays: galaxies: clusters --- acceleration of particles -- shock waves}

\section{Introduction}
\label{sec:Intro}

As a galaxy cluster grows, by accreting matter from its surrounding, a strong, collisionless, virial shock is thought to form at the so-called virial shock radius, $r_s$.
By analogy with supernova remnant (SNR) shocks, virial shocks too should accelerate charged particles to highly relativistic, $\gtrsim 10\TeV$ energies.
These particles, known as cosmic ray (CR) electrons (CREs) and ions (CRIs), should thus form a nearly flat, $E^2dN/dE\propto \const$ spectrum (equal energy per logarithmic CR energy bin), radiating a distinctive non-thermal signature which stands out at the extreme ends of the electromagnetic spectrum.

High-energy CREs cool rapidly, on timescales much shorter than the Hubble time $H^{-1}$, by Compton-scattering cosmic microwave-background (CMB) photons
\citep{LoebWaxman00, TotaniKitayama00, KeshetEtAl03}.
These up-scattered photons should then produce \gama-ray emission in a thin shell around the galaxy cluster, as anticipated analytically \citep{WaxmanLoeb00, TotaniKitayama00} and calibrated using cosmological simulations \citep{KeshetEtAl03, Miniati02}.
The projected \gama-ray signal typically shows an elliptic morphology, elongated towards the large-scale filaments feeding the cluster \citep{KeshetEtAl03,KeshetEtAl04}.
The same \gama-ray emitting CREs are also expected to generate an inverse-Compton ring in the optical band \citep{YamazakiLoeb15} and in hard X-rays \citep{KushnirWaxman10}, and a synchrotron ring in radio frequencies \citep{WaxmanLoeb00, KeshetEtAl04, KeshetEtAl04_SKA}.
Interestingly, the inverse-Compton signature in soft X-rays, which can stand out above the thermal background, was not previously explored in detail, to our knowledge.

By stacking Fermi Large Area Telescope (LAT; henceforth) data around 112 massive clusters, and by utilizing the predicted spatial and spectral dependence of the anticipated virial shock signal, the cumulative \gama-ray emission from many virial shocks was detected recently at a high ($>4.5\sigma$) significance \citep[][henceforth \Stack]{ReissEtAl17}.
The signal was found to be spectrally flat, with a photon spectral index $\alpha\equiv -d\ln N_\gamma/d\ln \epsilon=2.11_{-0.20}^{+0.16}$, and peaked upon radial binning around a radius $2.4R_{500}\simeq 1.5R_{200}$, in agreement with predictions.
Here, $N_\gamma$ and $\eps$ are the photon density and energy, and subscripts $\delta=200$ and $500$ designate an enclosed mass density $\delta$ times above the critical mass density of the Universe.
The signal indicates that the stacked shocks deposit on average $\xi_e\dot{m}\sim0.6\%$ of the thermal energy in CREs over a Hubble time.
As these results were obtained by radial binning, they sample only the radial component of the virial shocks, necessarily diluting the signal by picking up only those parts of the shocks favorably seen in such a projection.

It is interesting to study the signal from individual nearby clusters, where the signal may be picked up directly, without stacking.
The Coma cluster (Abell 1656), in particular, is one of the richest nearby clusters. With mass $M\sim 10^{15}M_\odot$, temperature $k_B T\sim 8\keV$, and richness class 2, it lies only $\sim 100\Mpc$ away \citep{GavazziEtAl09}, at a redshift $z\simeq 0.023$.
The cluster resides near the north Galactic pole (latitude $\sim88\dgr$), in a sky patch remarkably low on Galactic foreground.
These considerations, and indications for accretion as discussed below, render Coma exceptionally suitable for the search for virial shock signatures.

The virial radius of Coma, often defined as $R_v\simeq R_{200}\simeq 2.3\Mpc$ \citep{ReiprichBohringer02}\footnote{A wide range of $R_{200}$ estimates for Coma may be found in the literature, ranging from $1.8\Mpc$  \citep[self-similar extrapolation from the $R_{500}$ of][]{PiffarettiEtAl11}, to $2.1\Mpc$ \citep{GellerEtAl99}, $2.6\Mpc$ \citep{BrilenkovEtAl15}, and $2.8\Mpc$ \citep{KuboEtAl07}.}
, corresponds to an angular radius $\psi\simeq \psi_{200}\simeq 1\dgr.3$.
The cluster is somewhat elongated in the east--west direction, in coincidence with the western large scale structure (LSS) filament \citep{WestEtAl95} that connects it with the cluster Abell 1367. There is X-ray \citep{SimionescuEtAl13, UchidaEtAl16}, optical, weak lensing \citep{OkabeEtAl10,OkabeEtAl14}, radio \citep{BrownRudnick11}, and SZ \citep{PlanckComa12} evidence that the cluster is accreting clumpy matter and experiencing weak shocks towards the filament well within the virial radius, at $\psi\sim0.5\dgr$ angular radii.

An analysis \citep[][henceforth \Coma]{KeshetEtAl12_Coma} of a $\sim220\GeV$ VERITAS mosaic of Coma \citep{VERITAS12_Coma} found evidence for a large-scale, extended \gama-ray feature surrounding the cluster.
The apparent signal is best described as an elongated, thick, elliptical ring, with semiminor axis coincident with the cluster's virial radius, oriented toward the western LSS filament; the best fit was obtained for a
ratio $\zeta\equiv a/b\gtrsim2.5$ of semimajor axis $a$ to semiminor axis $b$.
The signal presents at a nominal $2.7\sigma$ confidence level, but there is substantial evidence supporting its presence and association with the virial shock.
This includes a higher, $5.1\sigma$ significance found when correcting for the observational and background-removal modes, indications that an extended signal was indeed removed by the background model, correlations with synchrotron and SZ tracers, good agreement ($3.7\sigma$) with the simulated \gama-ray ring of the cluster, and the absence of such extended signal tracers in VERITAS mosaics of other fields.
Interpreting the signal as a virial shock would imply $\xi_e\dot{m}\simeq1\%$, to within a systematic uncertainty factor of a few.

Other \gama-rays studies of Coma failed to detect a signal, largely because it is difficult to reach the combined high sensitivity, controlled foreground, good resolution, and high - yet not too high - photon energy, set by VERITAS.
For example, broad band, $>100\MeV$ analyses \citep{ZandanelAndo14, Prokhorov14} of LAT data found no excess emission from Coma, placing upper limits $\xi_i<15\%$ on CRI acceleration and $\xi_e<1\%$, and questioning spectrally flat emission matching the VERITAS signal.
However, at such low energies, the point spread function (PSF) is prohibitively large \citep{AtwoodEtAl13}, with $68\%$ ($95\%$) containment exceeding $5\dgr$, far beyond (exceeding $13\dgr$, an order of magnitude above) the $1.3\dgr$ virial radius.
LAT analyses of Coma thus rely on higher, $\gtrsim$GeV energies, where the photon statistics becomes increasingly challenging.
Moreover, the above upper limits are sensitive \citep{FermiComa16} to the assumed foreground, which is not accurately known at these energies, and to the morphology of the modeled signal.
Nevertheless, an extended LAT signal around Coma was eventually reported \citep{FermiComa16}, partly overlapping the virial radius.
This signal, still below the threshold needed to claim LAT detection, is consistent with the VERITAS signal when correcting for the larger extent of the latter (\Coma).

Here we use the signal identified in the VERITAS mosaic, to search for the counterparts of such inverse-Compton emission from the virial shock in other bands.
In \S\ref{sec:Theory} we study the expected energy-dependent morphology of the virial shock signal, in galaxy clusters and groups in general, and in Coma in particular.
We point out that at high, \gama-ray energies, the signal is expected to be spatially very narrow, due to the fast CRE cooling.
Furthermore, the ring in Coma may be even thinner than expected in a spherically (or triaxially) symmetric shock, as the distribution of LSS around Coma is approximately confined to a plane perpendicular to the line of sight, and such a planar configuration may manifest also in the accretion through the virial shock.
This does not contradict the thick VERITAS ring, which may have been considerably broadened by instrumental effects (on-region integration over a $0\dgr.4$ diameter region, ring background subtraction, and a $0\dgr.5$ wobble).

The LAT data, which is marginally sensitive to a signal at the level inferred from VERITAS, is analyzed in \S\ref{sec:LAT}.
We reproduce previously reported, LAT-based upper limits, in particular for a thick ring matching the VERITAS mosaic.
However, allowing for a thin signature, we find a high significance, elongated ring at the anticipated shock position and morphology.
We then analyze soft X-rays from ROSAT, where the virial shock signature can surface above the thermal signal, in \S\ref{sec:ROSAT}.
Here too, we find the expected signature, in the form of a thicker, smaller scale, elongated ring, as anticipated from the evolution of the low-energy CREs accelerated by the shock.
We then combine the signals from VERITAS, LAT, and ROSAT, and measure the overall spectrum, in \S\ref{sec:Combined}.
The three bands show a comparable logarithmic brightness, indicating an injected CRE spectral index $p\equiv - d\ln N_e/d\ln E\simeq2.0\till2.2$, consistent with a strong virial shock, and $\xi_e\dot{m}\simeq 0.3\%$.
Finally, the results are summarized and discussed in \S\ref{sec:Discussion}.

We adopt a flat $\Lambda$CDM cosmological model with a Hubble constant $H_0=70\km\se^{-1}\Mpc^{-1}$ and a mass fraction $\Omega_m=0.3$.
Assuming a $76\%$ hydrogen mass fraction gives a mean particle mass $\mass\simeq 0.59m_p$.
Confidence intervals quoted are $68\%$ for one parameter; multi-parameter  intervals are specified when used.
The results are primarily quantified in terms of an overdensity $\delta=500$; in Coma, $M_{500}=4.3\times 10^{14}M_\odot$, $R_{500}=1.14\Mpc$, and $\theta_{500}=0\dgr.68$ \citep{PiffarettiEtAl11}.
Accordingly, we define a normalized angular distance $\tau\equiv \theta/\theta_{500}$ from the center (defined as the X-ray peak) of Coma; $\theta$ (and $\tau$) is subsequently generalized to the (dimensionless) semiminor axis of an ellipse.

\section{Energy-dependent virial signature}
\label{sec:Theory}

\subsection{CRE evolution}

CREs of high energy $E$ cool rapidly by Compton up-scattering CMB photons, leading to a radiative signature in the form of a thin ring at photon energy $\eps\propto E^2$.
At low energies, the CRE cooling time $t_{cool}\propto E^{-1}\propto \epsilon^{-1/2}$ becomes long, so these CREs are able to propagate farther from the shock before radiating away their energy.
The resulting broadening of the virial shock signal with decreasing photon energy can be attributed to several effects, in particular the CRE (i) advection downstream, toward the center of the cluster; (ii) diffusion downstream; and (iii) escape upstream. Of these three processes, only downstream advection is reasonably well constrained on theoretical grounds.

The signal is not only broadened by the evolution of the CRE distribution, but is also distorted by it.
As CREs are advected deeper into the cluster, adiabatic compression raises both their density and their energy.
This locally boosts the brightness of the radiative signature, and shifts it toward higher energies.

If CRE diffusion were sufficiently strong, at low energies the virial shock signal would be greatly broadened, encompassing the entire cluster.
In addition, such diffusion would stem the adiabatic compression of the CREs.
The resulting smooth, faint radiative signal would be difficult to distinguish from the foreground, and may eventually be rendered undetectable.
As we show below in \S\ref{sec:ROSAT}, we do identify a localized signal inward of the virial shock even in soft X-rays.
Therefore, diffusion cannot be too strong.
In the following, we thus neglect CRE diffusion, and revisit it by imposing an upper limit on the diffusion function in \S\ref{sec:Discussion}.

If both downstream advection and upstream escape are sufficiently strong, one may see two signals arising from low energy CREs.
The first would be an inner, compression-enhanced signal, peaked at $\sim r_{cool}<r_s$, associated with cooling-limited inward advection.
The second would be an outer signal, peaked outside the shock radius $r_s$, associated with the escaping CREs.
The present data lack the sensitivity needed to detect the latter, putative, upstream component, so here we analyze only the former, inner signal.
The following discussion thus takes into account downstream advection only.

\subsection{Downstream advection}

To compute the radiative signature of the cooling, advected, CREs, one needs to evolve the CRE distribution.
This, in turn, requires an estimate of the time that elapsed since each mass element crossed the shock.
One way to do so is to approximate the accretion parameter $\mdot\equiv \dot{M}/(MH)\simeq \dot{M}_{bar}/(M_{bar}H)$ as constant throughout the evolution of the cluster, and to adopt some mass-radius relation such as the linear, $M_{bar}(r)\propto r$ profile of an isothermal sphere.
Here, subscript $bar$ designates baryons.
Then the radius of a shell that crossed the shock at an earlier time, $\Delta t>0$ ago, is given by
\begin{equation}
r(\Delta t) \simeq r_s e^{-f_m\mdot H_0 \Delta t} \coma
\end{equation}
where for simplicity we approximated the Hubble parameter by its present-day value, $H\simeq H_0$, and similarly took $r_s(\Delta t)\simeq r_s$.
The dimensionless factor $f_m$ accounts for deviations from the linear $M_{bar}\propto r$ relation, from constant $\dot{m}$, $H$ and $r_s$, and from a spherical geometry.

Consider an injected power-law CRE spectrum of index $p$, and in particular CREs that emit photons of typical energy $\eps\equiv \eps_{kev}\keV$.
The emissivity of these CREs, at times much shorter than their cooling time,
\begin{equation}
t_{cool} \simeq \frac{3m_e c}{4u_{cmb}\sigma_T}\sqrt{\frac{3k_B T_{cmb}}{\eps}} \simeq 2\eps_{keV}^{-1/2}\Gyr \coma
\end{equation}
scales as $j_\eps\propto \eps^{-(p-1)/2}$.
Here, $m_e$ is the electron mass, $c$ is the speed of light, $\sigma_T$ is the Thompson cross section, $T_{cmb}$ and $u_{cmb}$ are the CMB temperature and energy density, and $k_B$ is the Boltzmann constant.

For a nearly flat, $p\simeq 2$ spectrum, we may approximate the emission from each CRE as constant during its cooling time, so its time-integrated contribution to the brightness becomes $J_\eps \propto t_{cool}j_\eps\propto\eps^{-p/2}$.
The brightness observed at a projected normalized distance $\varrhoNorm\equiv \varrho/r_s$ from the center of the cluster, where $\varrho\equiv r_\perp$ is the distance in the plane of the sky, is then found by the line of sight integration through a shell of radial range $r$ given by $r_{cool}\equiv r(\Delta t= t_{cool})<r<r_s$.

We first consider a model in which the CREs are injected uniformly across a spherical shock.
This model, referred to as the shell model, is discussed in \S\ref{sec:ShellModel}.
An alternative, planar model, which may be more appropriate in the case of Coma, is motivated and discussed in \S\ref{sec:ExpectedComa}.
Both models are subsequently generalized for a triaxial shock surface in \S\ref{sec:EllipticModels}.

\subsection{Shell model}
\label{sec:ShellModel}

In the shell model, integration over the radiating layer of CREs is equivalent to taking the difference between the radiation due to the volume inside $r_s$, and the radiation from the volume inside $r_{cool}$.
This yields the logarithmic brightness
\begin{equation} \label{eq:JModel}
\eps J_\eps(\varrhoNorm) = \int \eps j_\eps \, dl = A(\eps) [B(\varrhoNorm;1)-B(\varrhoNorm;\varrhoNorm_{cool})] \coma
\end{equation}
where $\varrhoNorm_{cool}\equiv r_{cool}/r_s$, and $B(\varrhoNorm;\varrhoNorm_{max})$ is the dimensionless brightness at $\varrhoNorm$ due to a globe of radius $\varrhoNorm_{max}$, assuming unit emissivity at radius $\varrhoNorm_{s}=1$.
Here we defined the ($\varrho$-independent) normalization
\begin{equation}\label{eq:ADef}
A(\eps)
\equiv \epsilon j_\epsilon(r_s) r_s
\propto \eps^{\frac{3-p}{2}}
\end{equation}
as the (non-cooled) emissivity at the shock surface, weighted by the shock radius.

In the absence of brightening due to CRE compression, integration over the homogeneous volume of the shell gives the geometrical factor
\begin{equation}
B(\varrhoNorm;\varrhoNorm_0)= B_0\equiv \left(\varrhoNorm_0^2-\varrhoNorm^2\right)^{1/2}\Theta(\varrhoNorm_0-\varrhoNorm) \fin
\end{equation}
Here, $\Theta$ is the Heaviside step function.
Incorporating a power-law compression of the CRE energy, of the form $u_{cre}\propto r^{-q}$, gives instead
\begin{equation} \label{eq:VSB}
B(\varrhoNorm;\varrhoNorm_0) = B_0 \varrhoNorm^{-q} {}_2F_1\left(\frac{1}{2},\frac{q}{2};\frac{3}{2};1-\frac{\varrhoNorm_0^2}{\varrhoNorm^2}\right) \coma
\end{equation}
where ${}_2F_1$ is the hypergeometric function.
One may crudely approximate the compression using the projected radius,
\begin{equation} \label{eq:VSB0}
B(\varrhoNorm;\varrhoNorm_0) \simeq B_0 \mbox{Max}\left(\varrhoNorm,\varrhoNorm_{cool}\right)^{-q}  \coma
\end{equation}
where the brightening factor is not allowed to exceed $\varrhoNorm_{cool}^{-q}$ in order to avoid excessive compression and nonphysical divergence at small radii.
The approximate Eq.~(\ref{eq:VSB0}), although useful for $\varrhoNorm\gtrsim1/2$, is not utilized in what follows.

To estimate the compression index $q$, we first consider an isothermal sphere distribution, where the gas number density scales as $n\propto r^{-2}$, and assume for simplicity that all mass elements were shocked to the same density and temperature \citep[\eg][]{KushnirWaxman09}.
Hence, $u_{cr}\propto n^{4/3}\propto r^{-8/3}$, such that q=8/3, where we assumed adiabatic compression of CRs with an adiabatic index $\Gamma=4/3$.
Such a model is consistent with the $n\propto r^{-3\beta}\sim r^{-2}$ profile of Coma's $\beta$-model when extrapolated to large radii.
It can be directly generalized for the steeper profile expected at the cluster periphery.
For example, the steep, $n\propto r^{-4}$ profile found at large radii in the Hernquist model \citep{Hernquist90} yields $q=16/3$.
Note that due to the compression, the energy of each CRE increases as $E\propto n^{1/3}\propto r^{-2/3}$ for an isothermal sphere, and as $E\propto r^{-4/3}$ for an $n\propto r^{-4}$ profile.

\begin{figure*}
	\centerline{
        \epsfxsize=9cm \epsfbox{\myfig{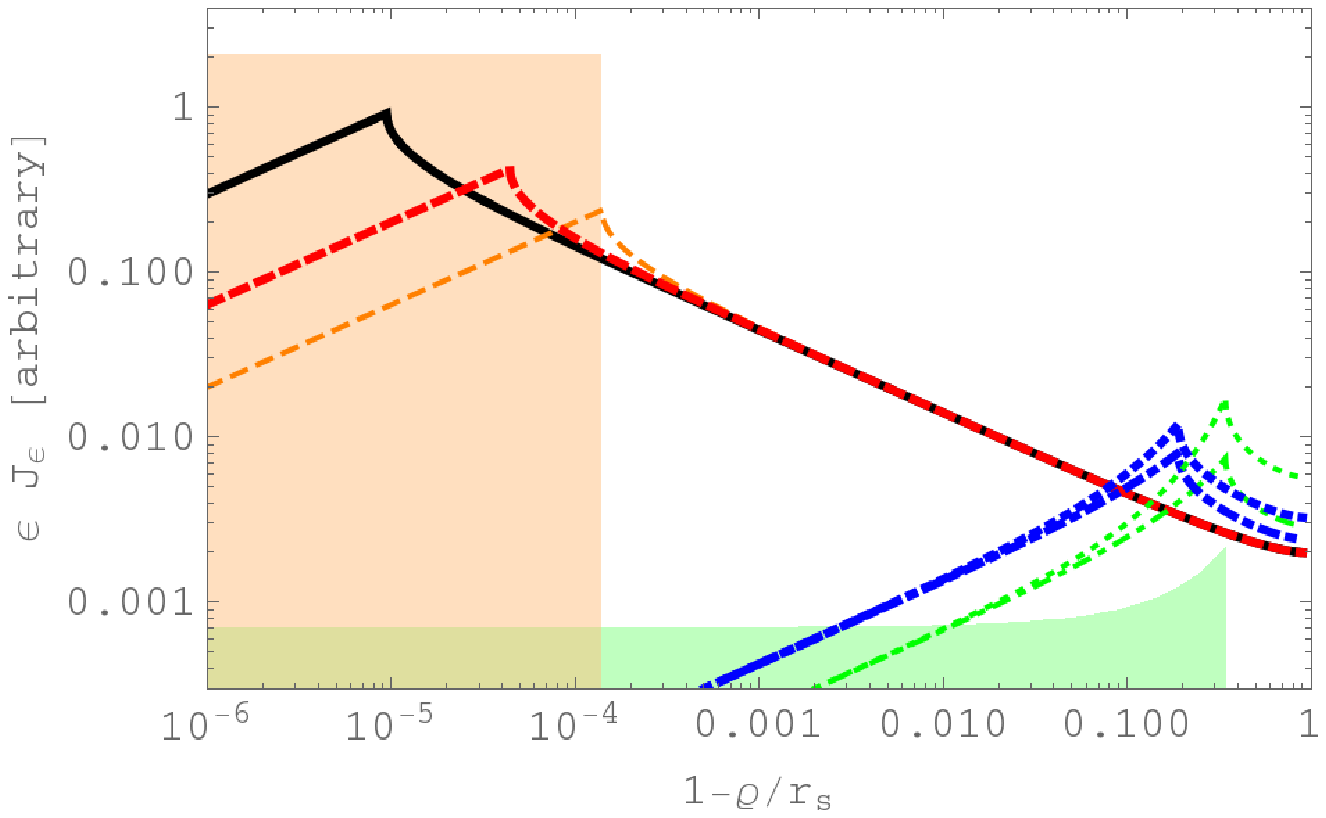}}
        \epsfxsize=8.8cm \epsfbox{\myfig{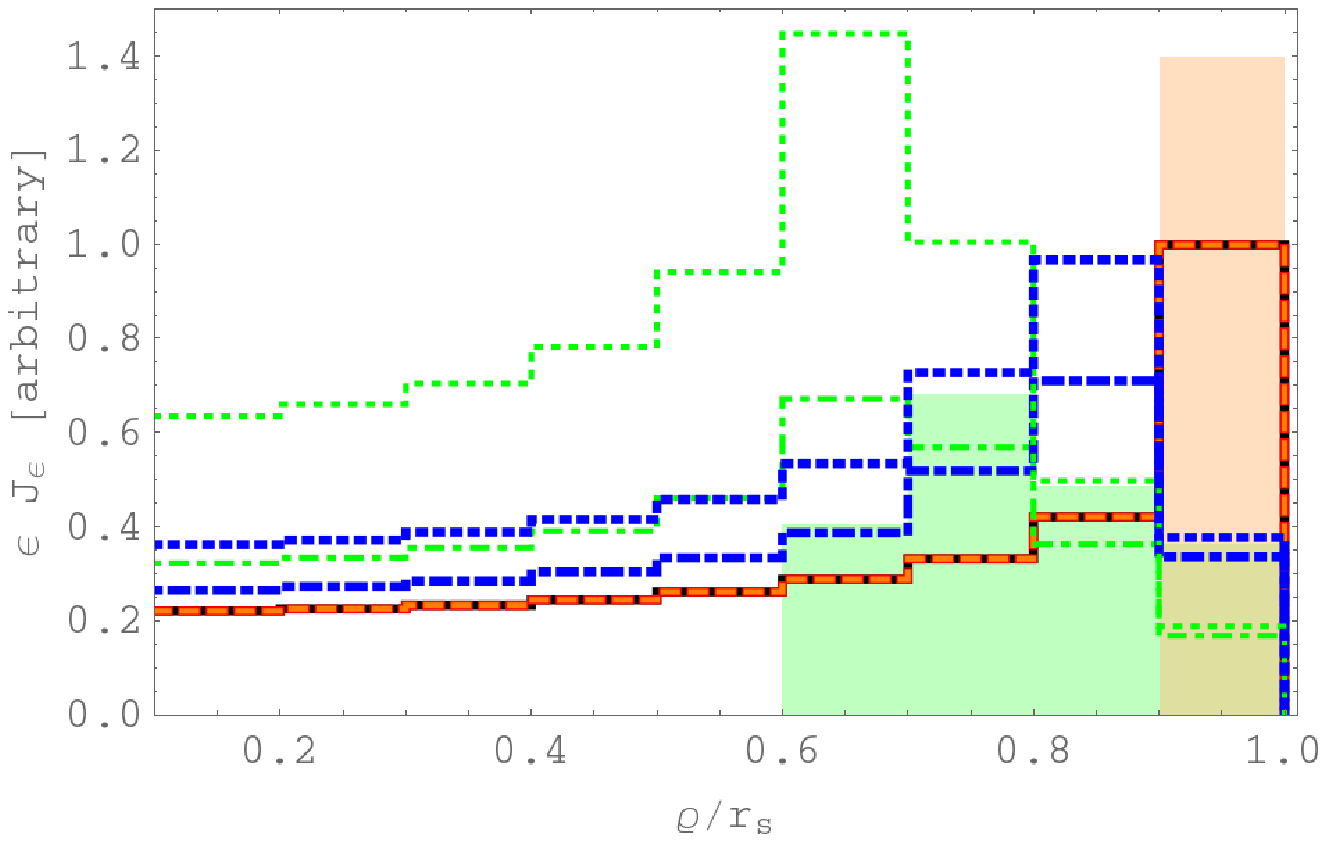}}
    }
\caption{\label{fig:VSsingature}
Brightness of inverse-Compton emission from shock-accelerated CREs with a flat, $p=2$ spectrum, shown both unbinned (left panel, as a function of normalized, projected distance $1-\varrho/r_s$ from the shock) and binned (right, as a function of normalized radius $\varrho/r_s$), both for the shell model (curves; Eqs.~\ref{eq:JModel}--\ref{eq:ADef} and \ref{eq:VSB} unbinned, Eqs.~\ref{eq:JModelBinned}--\ref{eq:VSB2} binned) and for the planar model (shaded regions; Eqs. \ref{eq:JModelPlanar}--\ref{eq:APlanarDef} unbinned, Eq.~\ref{eq:VSBPlanar} binned).
The injected CREs are assumed to be advected downstream and adiabatically compressed, with $f_m\mdot=1$.
Results are shown for the energy bands of VERITAS ($\sim220\GeV$; solid black), \emph{Fermi}-LAT (dashed; thick red for $10\GeV$, thin orange for $1\GeV$), and ROSAT (thick blue for $0.44\keV$ in band R4, thin green for $0.11\keV$ in band R1; dot-dashed for $q=8/3$, dotted for $q=16/3$).
The planar model, in which injection is confined to the plane of the sky, is illustrated for the LAT and ROSAT R1 bands (shaded regions for $1\GeV$ and for $0.11\keV$ with $q=8/3$).
The right panel uses $10$ uniform bins, roughly corresponding to our nominal LAT and ROSAT analyses.
}
\end{figure*}

The resulting brightness profile from the virial shock is illustrated in Figure \ref{fig:VSsingature} for different energy bands.
For comparison with binned data, we also compute the signature binned onto projected radial annuli.
Integration over both the line of sight and the projected radius $\varrho$ in the bin $\varrho_i<\varrho<\varrho_{i+1}$ yields a mean bin brightness
\begin{equation} \label{eq:JModelBinned}
\langle \epsilon J_\epsilon \rangle = \frac{\epsilon \tilde{J}_\epsilon(\varrhoNorm_{i+1})-\epsilon \tilde{J}_\epsilon(\varrhoNorm_{i})}{\varrhoNorm_{i+1}^2-\varrhoNorm_{i}^2} \coma
\end{equation}
where $\epsilon \tilde{J}_\epsilon$ is given by the RHS of Eq.~(\ref{eq:JModel}), but with $B$ replaced by
\begin{equation} \label{eq:VSB2}
\tilde{B} = \frac{4B_0}{q-2}\left[\varrhoNorm^{2-q} {}_2F_1\left(\frac{1}{2},\frac{q-2}{2};\frac{3}{2};1-\frac{\varrhoNorm_0^2}{\varrhoNorm^2}\right)-\varrhoNorm_0^{2-q}\right] \fin
\end{equation}
The binned profile is illustrated in the right panel of Figure \ref{fig:VSsingature}, for the same energy bands.

\subsection{Expected signal in Coma: planar model}
\label{sec:ExpectedComa}

When focusing on a specific cluster such as Coma, rather than stacking data over many clusters, it is advantageous to investigate the large-scale environment of the cluster. This environment is likely to affect the accretion pattern of the cluster, which is directly reflected in the \gama-rays from the virial shock, and, with some time delay, also in the nonthermal X-rays.
This includes for example a brightening in localized regions with an enhanced accretion rate, leading to bright spots along the virial shock surface \citep{KeshetEtAl03}.

\begin{figure}[b]
	\centerline{\epsfxsize=9cm \epsfbox{\myfig{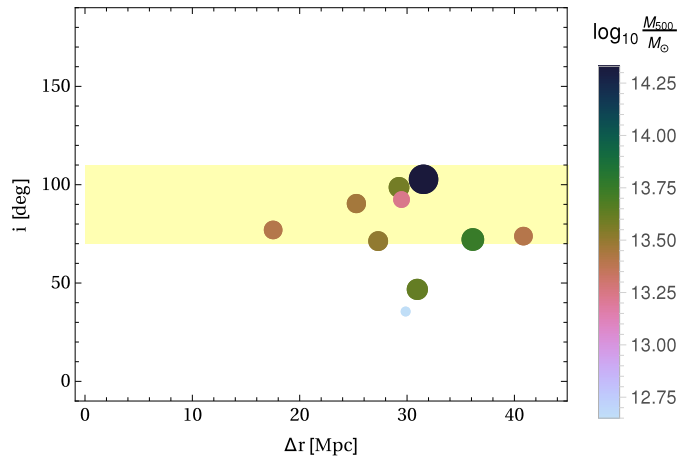}}}
\caption{\label{fig:ComaEnv}
Neighboring galaxy clusters and galaxy groups within $50\Mpc$ of Coma, plotted (disks with colors and sizes corresponding to the cluster or group mass $M_{500}$; see colorbar) in the plane of proper distance $\Delta r$ from Coma vs. inclination $i$ with respect to the Coma line of sight. Most LSS objects approximately lie within a plane perpendicular to the line of sight, at inclinations $|i-90\dgr|<20\dgr$
%$70\dgr<i<110\dgr$
(dashed yellow region). The most massive cluster shown (largest, darkest disk) is A1367.
}
\end{figure}

We thus examine the galaxy clusters and groups in the nearby, $\Delta r<50\Mpc$ vicinity of Coma, using the Meta-Catalog of X-ray Clusters \citep[MCXC;][]{PiffarettiEtAl11}.
Let $i$ be the inclination of a LSS object with respect to Coma, defined as the angle between the Coma line of sight and the line connecting the object and Coma.
As Figure \ref{fig:ComaEnv} shows, all the large scale structure in the vicinity of Coma lies at inclination angles $i$ nearly perpendicular to the line of sight, most of it within the range $|i-90\dgr|<20\dgr$.
These objects are found at a typical proper distance $\Delta r\sim30\Mpc$ from Coma.
Massive clusters $A1185$ and $A1177$ have smaller, $i\simeq 50\dgr$ inclinations; however, at $\Delta r\sim60\Mpc$ from Coma, these two clusters cannot be considered as part of its close environment.

Assuming that the accretion rate through the Coma virial shock correlates spatially with the distribution of surrounding galaxy clusters and groups, one would therefore expect the virial emission to be particularly strong in the plane of the sky,
\ie in the vicinity of the plane cutting through Coma perpendicular to the line of sight.
The shell model, integrating over a spherically symmetric shell and shown as curves in Figure \ref{fig:VSsingature}, would then overestimate the signal at small radii.
For preferentially planar emission, the signal from the virial shock would drop rapidly as the projected radius $\varrho$ decreases inward.
Indeed, in the limit of emission confined to a plane, the signal would drop to zero within one cooling distance from the projected shock radius, at $\varrho\simeq r_{cool}$.

Consider the limit of a thin, planar emission layer of width $\Delta h$, henceforth referred to as the planar model.
The brightness here would still be given formally by Eq.~(\ref{eq:JModel}), but with the normalization $A$ of Eq.~(\ref{eq:ADef}) replaced by
\begin{equation} \label{eq:APlanarDef}
\tilde{A}(\eps)\equiv (\epsilon j_\epsilon)_{r_s}\Delta h
\coma
\end{equation}
and Eq.~(\ref{eq:VSB}) replaced by
\begin{equation}\label{eq:JModelPlanar}
B(\varrhoNorm;\varrhoNorm_0) =  \varrhoNorm^{-q}\Theta(\varrhoNorm_0-\varrhoNorm) \fin
\end{equation}
The binned brightness would similarly be given by Eq.~(\ref{eq:JModelBinned}), but using $\tilde{A}$ instead of $A$, and with Eq.~(\ref{eq:VSB2}) replaced by
\begin{equation}\label{eq:VSBPlanar}
\tilde{B} = \frac{2}{q-2} \left(\varrhoNorm^{2-q}-\varrhoNorm_0^{2-q}\right)
\Theta(\varrhoNorm_0-\varrhoNorm) \fin
\end{equation}

The brightness profile in the planar model is illustrated as shaded regions in Figure \ref{fig:VSsingature}, for the LAT and ROSAT energy bands.
The sharp drop in brightness inside $\varrho_{cool}$ reflects our assumption of a well-defined cooling time; in practice, this drop will be somewhat smoothed.

\subsection{Elliptic generalization}
\label{sec:EllipticModels}

When studying an individual cluster, one should also consider the morphology of the virial shock, in particular its projected elongation (\ie the ratio $\zeta$ of semimajor axis to semiminor axis) on the sky.
The dark matter halos of galaxy clusters are thought to be nearly prolate, especially when the cluster is unrelaxed \citep{LemzeEtAl12}, with a typical (three dimensional) elongation $\zeta\simeq 2$ \citep[\eg][]{GroenerGoldberg14} that becomes larger for more massive clusters, reaching an average $\zeta\simeq 2.2$ at the high mass end \citep{DespaliEtAl14}.
The distribution of SDSS galaxies around Coma (\Coma) suggests an even more elongated structure, around $\zeta\simeq 2.5$.
Indeed, $\zeta\simeq 2.5$ is the minimal elongation for which the VERITAS signal reaches peak significance (\Coma); however, this signal remains strong for even more elongated templates, as large as the VERITAS mosaic allows ($2.5\lesssim\zeta\lesssim4$).
In conclusions, elongation values in the range $2\lesssim\zeta\lesssim3$ are expected in the virial shock of Coma.

Next, consider the orientation of the virial ring's elongation.
Based on numerical simulations, the major axis is typically expected to point in the direction of the main connecting galaxy filament, presumably related to the most massive LSS neighbor.
In the case of Coma, this object is A1367 (shown in Figure \ref{fig:ComaEnv} as the largest, darkest disk). With $M_{500}\simeq 2.1\times 10^{14}M_\odot$, this cluster is less massive than Coma, but $\sim 5$ times more massive than Coma's second most massive neighbor, NGC 4104.
The expected orientation of the virial ring, based on the distribution of SDSS galaxies toward A1367, is indeed consistent with the preliminary signal in the VERITAS mosaic, both found (\Coma) to peak at major axis angles $-10\dgr<\phi\lesssim 0\dgr$, where $\phi$ is the azimuthal angle in equatorial coordinates (such that $\phi=0\dgr$ points due west).

The preceding discussion, invoking the distribution of LSS around Coma and the preliminary signal from VERITAS, motivates a search for an elliptic virial ring signature, of typical $\zeta\sim2.5$ elongation with the major axis in the east--west direction, and suggests that the emission along the virial shock surface would be preferentially strong in the plane of the sky.
According to the VERITAS data analysis (see figure 2 in {\Coma}), due to the elongation of the ring, searching for a purely circular feature is not expected to show any virial signal in Coma.
Instead, one should consider an elliptical template.
A simple approach is to adopt the spherical signature derived above, as illustrated in Figure \ref{fig:VSsingature}, and simply stretch it uniformly along the major axis.

We therefore redefine (henceforth) $\varrho$ as a projected elliptical radial coordinate, $\varrho\equiv (\varrho_b^2+\varrho_a^2/\zeta^2)^{1/2}$, where $\varrho_b$ and $\varrho_a$ are the projected coordinates along the minor and major axes, respectively.
For comparison with other clusters and with the stacking analysis of {\Stack}, it is convenient to normalize scales with respect to  overdensity parameters such as those evaluated at $\delta_{500}$.
A dimensionless elliptical coordinate $\tau\equiv \varrho/R_{500}$ is thus defined, by normalizing $\varrho$ with respect to $R_{500}$.
Equivalently, in terms of the angular separation  $\theta$ from the center of Coma, we may define $\tau\equiv \theta/\theta_{500}$, where $\theta\equiv (\theta_b^2+\theta_a^2/\zeta^2)^{1/2}$.
The coordinate $\tau$ may be regarded as the normalized semiminor axis of an ellipse.
For a spherical shock, our definition of $\tau$ coincides with that of  {\Stack}.
The VERITAS analysis suggests that the shock should lie near $\tau\simeq 2$.

\section{Fermi-LAT analysis}
\label{sec:LAT}

\subsection{Data preparation and analysis}

\begin{figure*}
	\centerline{
        \begin{overpic}[width=8.5cm, trim=0 0cm 0 0cm]{\myfig{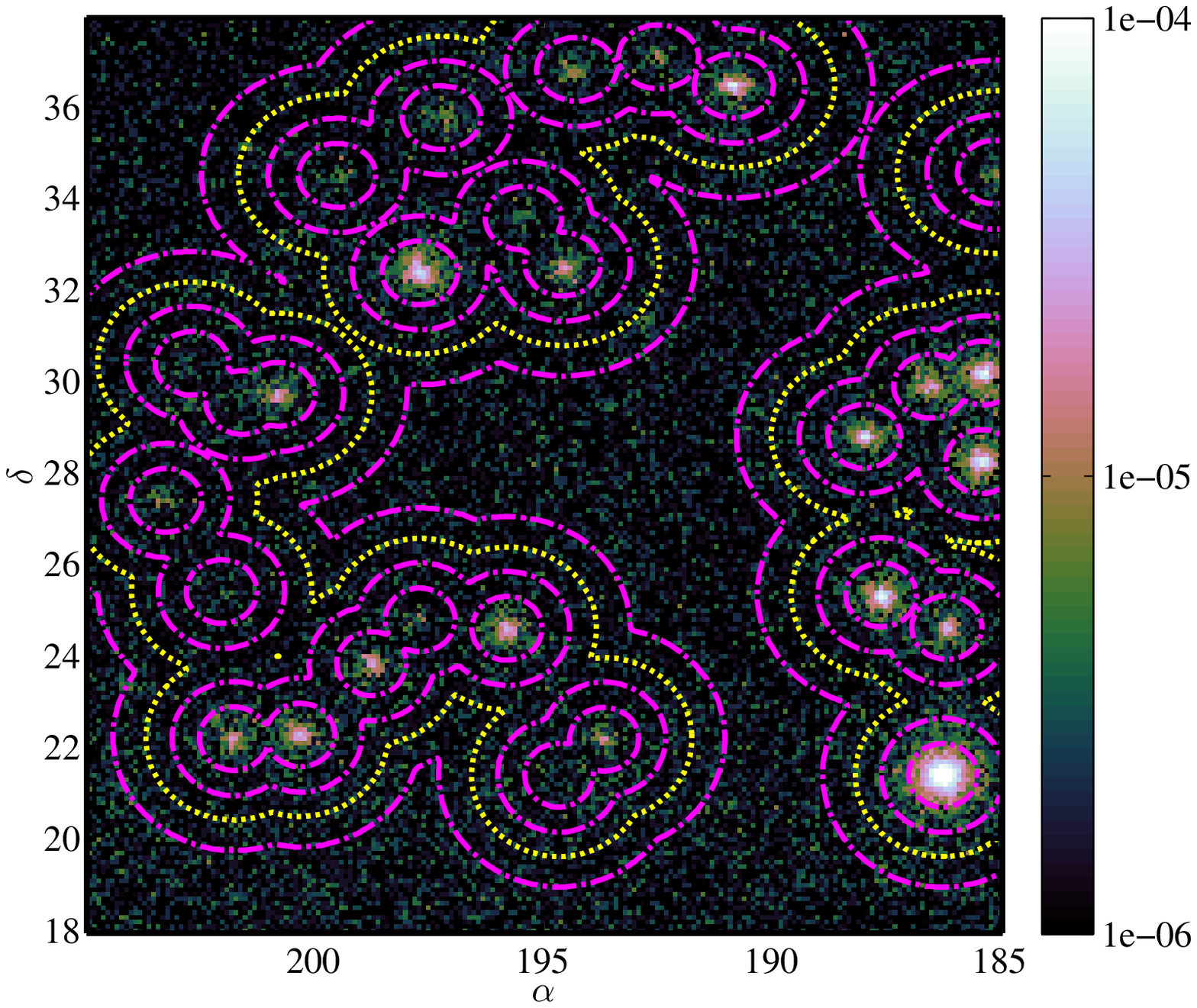}}
        \put (8,78) {\normalsize \textcolor{white}{(a)}}
        \end{overpic}
        \begin{overpic}[width=8.5cm, trim=0 0cm 0 0cm]{\myfig{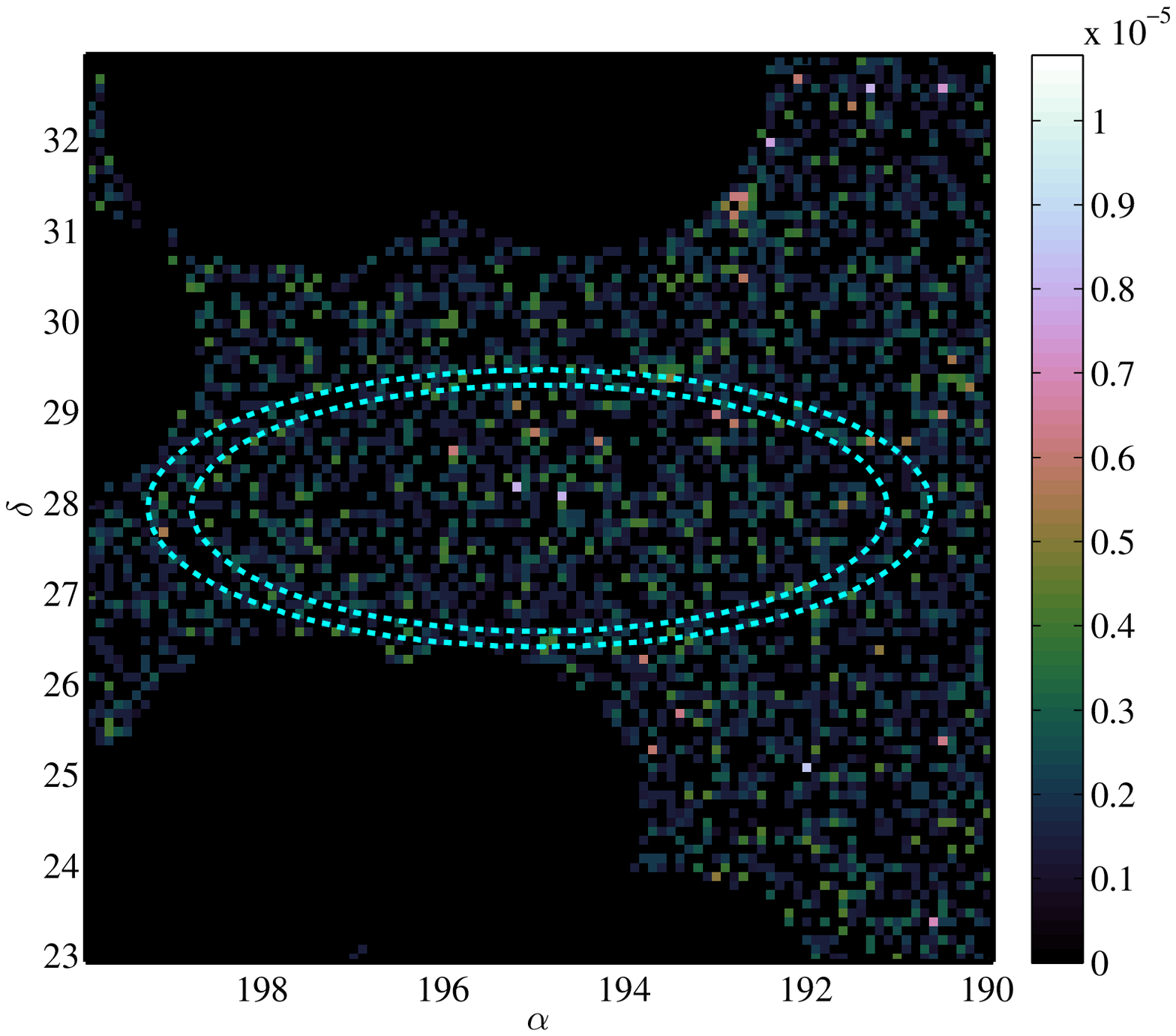}}
        \put (8,78) {\normalsize \textcolor{white}{(b)}}
        \end{overpic}
    }
    \caption{\label{fig:ComaFermiMap}%
	LAT $>1\GeV$ brightness \citep[$\log_{10}J_\eps\mbox{[s}^{-1} \cm^{-2} \sr^{-1}\mbox{]}$ cubehelix colorbar;][henceforth]{Green11_Cubehelix} map centered upon Coma  in a CAR projection and equatorial coordinates.
    The $95\%$ containment angles around 3FGL point sources at 1, 3.2 and $10\GeV$ energies are superimposed (dot-dashed magenta contours) in panel (a). The $90\%$ containment angle at 1 GeV is shown (dotted yellow) in panel (a) and masked in the zoomed-in panel (b).
    The $2<\tau<2.25$ elliptic bin in our nominal, east--west ($\phi=0\dgr$) elongated ($\zeta=2.5$) ring is highlighted in panel (b) (dashed cyan), as a guide to the eye.
    The signal in this bin is not sufficiently brighter than the foreground to be easily discernable by eye.
	}
\end{figure*}

We use the archival, $\sim8$ year, Pass-8 LAT data from the Fermi Science Support Center (FSSC)\footnote{\texttt{http://fermi.gsfc.nasa.gov/ssc}}, and the Fermi Science Tools (version \texttt{v10r0p5}).
Pre-generated weekly all-sky files are used, spanning weeks $9\till422$ for a total of 414 weeks ($7.9\yr$), with SOURCE class photon events.
A zenith angle cut of $90\dgr$ is applied to avoid CR-generated $\gamma$-rays originating from the Earth's atmospheric limb, according to the appropriate
FSSC Data Preparation recommendations.
Good time intervals are identified using the recommended selection expression \texttt{(DATA\_QUAL==1) and (LAT\_CONGIF==1)}.

Sky maps are discretized using a HEALPix scheme \citep{GorskiEtAl05} of order $N_{hp}=10$, providing a mean $\sim 0.057\dgr$ pixel separation.
This is sufficient for analyzing the anticipated $\sim1\dgr.5 \times 4\dgr$ virial shock in the Coma cluster, and is smaller than the $0\dgr.2$ high-energy PSF of the LAT \citep[$68\%$ containment angle at $E\gtrsim10$ GeV;][]{AtwoodEtAl13}.

Event energies are logarithmically binned onto $N_\epsilon=4$ energy bands in the (1--100) GeV range.
Unlike the all-sky analysis of {\Stack}, here we do not rely on the highest energy band, due to the poor statistics in the small region around a single cluster; we subsequently confirm that incorporating this band does not modify our conclusions.
Point source contamination is minimized by masking pixels within the $95\%$ containment area of each point source in the LAT 4-year point source catalog \citep[3FGL;][]{FermiPSC}.
The LAT data around Coma and the masking of point sources are presented in Figure \ref{fig:ComaFermiMap}, on both large ($20\dgr$) and small ($10\dgr$) scales.

The foreground, after point sources were masked, varies mainly on scales much larger than the sub-degree width of the anticipated signal.
Therefore, this remaining foreground can be accurately approximated using a polynomial fit on large scales.
We thus consider a large, $\tau<\tau_{max}\equiv 8$ (equivalently, $\theta\lesssim 5\dgr.4$) elliptical disk region around Coma, and fit the corresponding LAT data by an order $N_f=4$ polynomial in the normalized angular coordinates $\tau_x$ and $\tau_y$.
To minimize the effect of the central diffuse signal (discussed below), we use only the $\tau>\tau_{min}=1.5$ ($\theta\gtrsim1\dgr$) data for foreground estimation.
The foreground is evaluated separately in each of the three energy bands.

Sensitivity tests, presented in part in {\Stack} for the purpose of stacking analyses, and established here for the analysis of Coma, indicate that the results do not strongly depend on the precise choice of analysis variants and parameters, as discussed in \S\ref{sec:LATRobustness} below.

We bin the LAT data into concentric elliptical rings about the center of Coma, assuming a ring morphology defined by an elongation $\zeta$ and a major axis  orientation $\phi$.
For each photon energy band $\epsilon$, and each radial bin centered on $\tau$ with width $\Delta \tau$, we define the excess emission $\Delta n\equiv n-f$ as the difference between the number $n$ of detected photons, and the number $f$ of photons estimated from the fitted foreground.
The significance of the excess emission in a given energy band $\epsilon$ and radial bin $\tau$ can then be estimated, assuming Poisson statistics with $f\gg1$, as
\begin{equation} \label{eq:SingleBinSignificance}
\nu_{\sigma}(\epsilon,\tau) \simeq {\Delta n}/{\sqrt{f}} \fin
\end{equation}

\subsection{Ring signal}

\begin{figure}[b]
	\centerline{
    %\epsfxsize=8.5cm \epsfbox{\myfig{ComaFlux1.eps}}
    %\vspace{3cm}
	%\epsfxsize=8.5cm \epsfbox{\myfig{ComaFlux2.eps}}
    \includegraphics[width=10cm,keepaspectratio=true]{\myfig{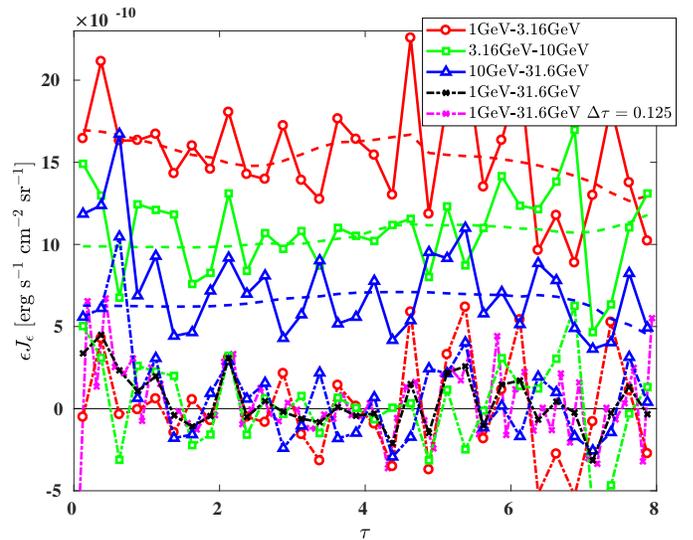}}
	}
\caption{\label{fig:LAT_flux}
LAT logarithmic energy flux in our nominal, east--west elongated, elliptical, concentric bins about the center of Coma.
The binned flux is shown in each of the three energy bands (symbols with solid lines to guide the eye; see legend), as a function of the normalized semiminor axis $\tau$, with bin size $\Delta\tau=0.25$.
The estimated foreground in each energy band (dashed curves) is based on a fourth-order polynomial fit.
The excess emission, shown for each band (lower symbols, dash-dotted lines) and for an average over the three bands (black x-marks), suggests a signals in the $2.0<\tau<2.25$ bin, as well as some diffuse emission near the center ($\tau\lesssim 1$).
Also shown is the band-averaged excess emission with narrower, $\Delta\tau=0.125$ bins (magenta crosses, dash-dotted lines).
}
\end{figure}

Figure \ref{fig:LAT_flux} shows the fluxes corresponding to $n$ (solid lines), $f$ (dashed) and $n-f$ (dash-dotted), for the three energy bands used.
Here, we adopt the nominal ring parameters inferred in \S\ref{sec:ExpectedComa}, namely $\zeta=2.5$ and $\phi=0\dgr$, and an angular bin size $\Delta \tau=0.25$, small enough to capture the anticipated thin \gama-ray ring.
The excess flux inferred from the three energy bands combined (black x-marks) shows a tentative signal in the $2.0<\tau<2.25$ bin.
This bin is highlighted by dashed contours in panel (b) of Figure \ref{fig:ComaFermiMap}.
This signal is resolved by splitting the bin further, as demonstrated in the figure for $\Delta \tau=0.125$ (magenta crosses).
The location of this signal, corresponding to the semiminor axis range $1\dgr.4\lesssim b\lesssim 1\dgr.5$, agrees with the location of the VERITAS signal ($1\dgr.0<b<1\dgr.6$).

\begin{figure}[h!]
	\centerline{\epsfxsize=8.5cm \epsfbox{\myfig{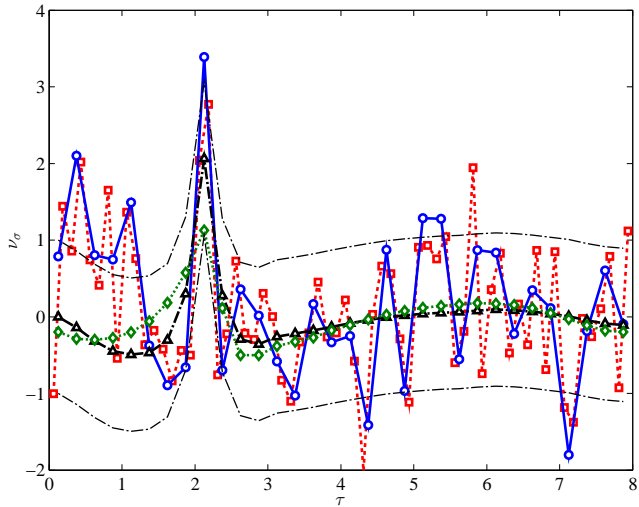}}}
    %\centerline{\includegraphics[angle=0,width=8.4cm,trim=0 30 0 0]{\myfig{ComaRing1.eps}}}
    \caption{\label{fig:LAT_significance}
	Co-added significance of LAT excess counts for our nominal binning. The significance of Eq.~(\ref{eq:SingleBinSignificance}) is shown for elliptic bins of width $\Delta\tau=0.25$  (circles, with solid blue line to guide the eye) and $\Delta\tau=0.125$ (squares; dashed red).
	Also shown are the results (with best fit parameters) of the shell model (diamonds; green dotted) and the planar model (triangles; black dash-dotted; with $1\sigma$ intervals as thin black dash-dotted curves).
	}
\end{figure}

The significance of the flux excess above the foreground is shown in Figure \ref{fig:LAT_significance} for the nominal ring morphology.
Some significant  diffuse excess can be seen in the central $\sim 1R_{500}$, the nature of which is beyond the scope of the present analysis, and is deferred to a future paper.
The narrow, elliptical ring-like signal in the $2.0<\tau<2.25$ bin presents at a $3.4\sigma$ confidence level for the nominal, $\Delta \tau=0.25$ bin width.
This excess can be resolved.
For narrower, $\Delta \tau=0.125$ binning, it is resolved into two, $2.8\sigma$ and $2.0\sigma$ confidence level, sub-bins.
The excess is more significant on the western side of the cluster.

\begin{figure}[h!]
	\centerline{\epsfxsize=9cm \epsfbox{\myfig{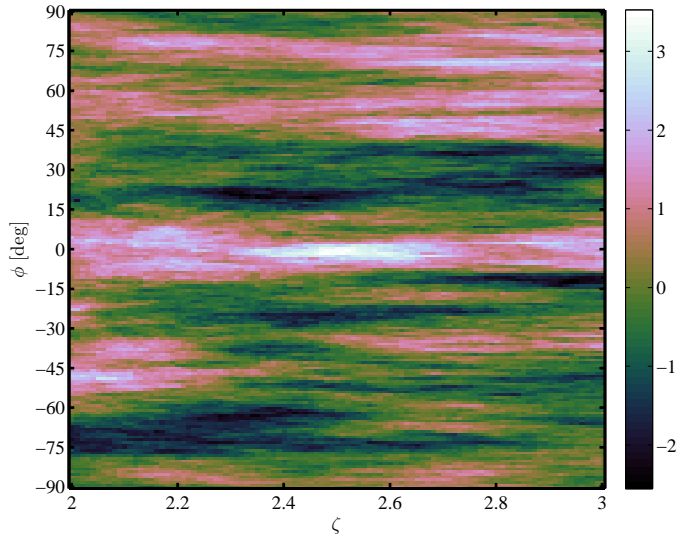}}}
    \caption{\label{fig:LAT_Scan}%
	Co-added significance of LAT excess counts in the $2.0<\tau<2.25$ elliptic bin, for different values of the  elongation $\zeta$ and orientation $\phi$ ring parameters.
	%using (a) $95\%$ masking of 3FGL sources and (b) $90\%$ masking.
	The maximal value
	%, in both cases,
	is obtained for $\zeta\simeq 2.5$ and $\phi\lesssim0\dgr$, in resemblance of the VERITAS signal.
	}
\end{figure}

Figures \ref{fig:LAT_flux} and \ref{fig:LAT_significance} pertain to the nominal, VERITAS-motivaed ring morphology, $\zeta=2.5$ and $\phi=0\dgr$.
Next, we examine different ring morphologies, by varying the values of $\zeta$ and of $\phi$. Figure \ref{fig:LAT_Scan} shows the significance of the $2.0<\tau<2.25$ signal for a wide range of plausible $\zeta$ and $\phi$ values.
It indicates that the nominal parameters approximately maximize the significance of the emission in this bin, in resemblance of the VERITAS signal.
The maximal bin significance is $3.6\sigma$, with  $\zeta=2.50_{-0.04}^{+0.07}$ and $\phi=-0\dgr.1_{-1\dgr.9}^{+0\dgr.4}$ encompassing the nominal parameters.

We find no such pronounced maximum in other $\tau$ bins in the relevant ($1<\tau<2.5$) range, for any $\{\zeta,\phi\}$ values.
We do find a broad, $\sim3.1\sigma$ maximum in the $1.75<\tau<2.0$ bin, but this corresponds to a $\phi\sim-10\dgr$ signal which partly overlaps with the same ring signature found, with a higher significance, in the $2.0<\tau<2.25$ bin.
Figure \ref{fig:LAT_Scan} pertains to the masking of $95\%$ containment around 3FGL sources, but we obtain nearly identical results for $90\%$ containment, indicating that the conclusions are independent of the masking pattern.

\begin{figure*}
	\centerline{
	    \begin{overpic}[width=9cm, trim=0 0cm 0 0cm]{\myfig{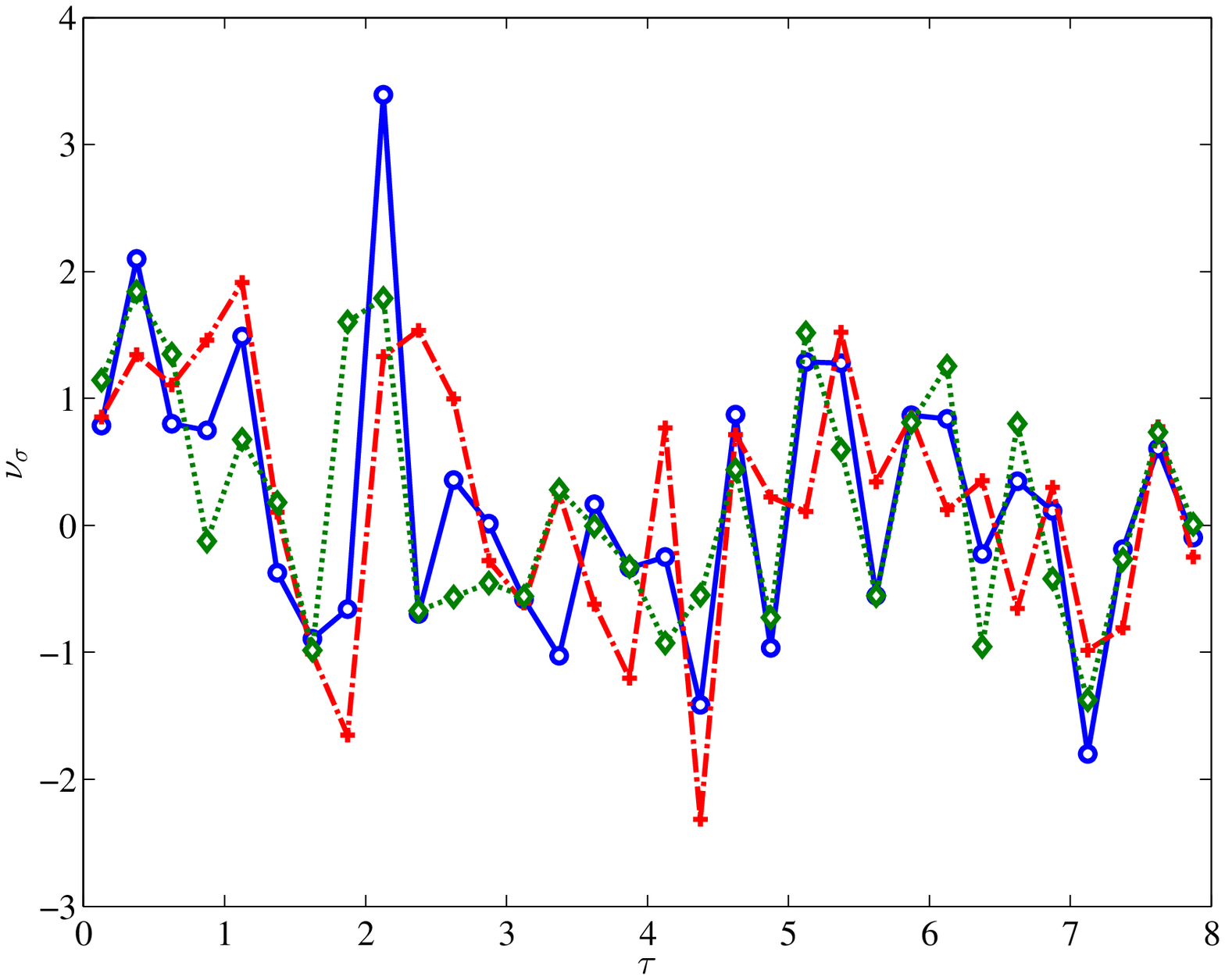}}
        \put (8,73) {\normalsize \textcolor{black}{(a)}}
        \end{overpic}
	     \begin{overpic}[angle=0,width=4cm,trim=0 0 0 0,clip]{\myfig{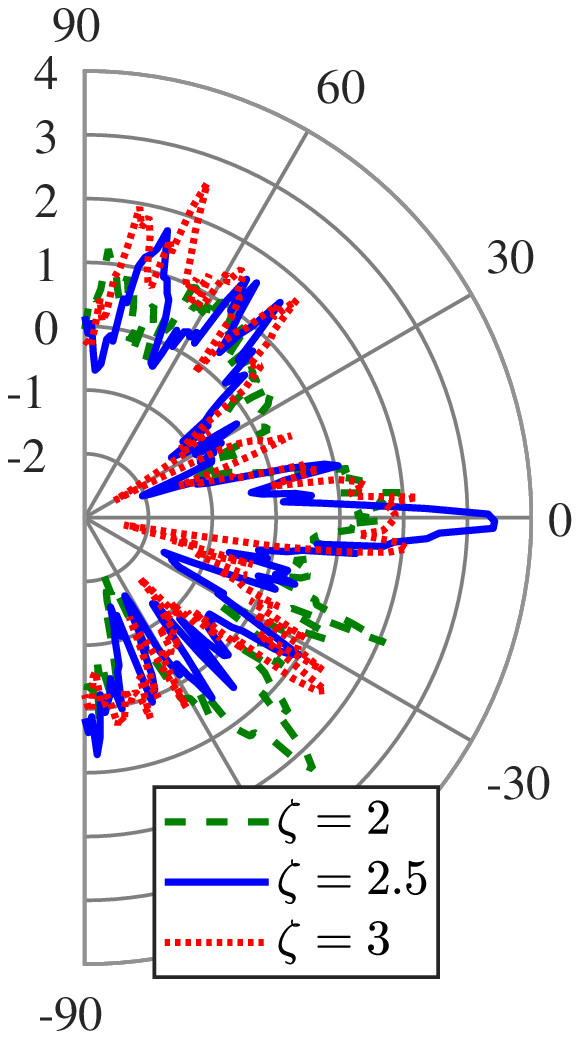}}
        \put (48,88) {\normalsize \textcolor{black}{(b)}}
        \end{overpic}
	}
    %\centerline{\includegraphics[angle=0,width=8.4cm,trim=0 30 0 0]{\myfig{ComaRing1.eps}}}
    \caption{\label{fig:LAT_Scans}%
	Illustrating the dependence of the LAT signal (co-added significance of excess counts) upon ring morphology parameters.
	Curves are shown for $\zeta=2.5$ (solid blue), $2$ (dotted green), and $3$ (dot-dashed red).
	In panel (a), the ring orientation is fixed at $\phi=0\dgr$.
	In panel (b), the elliptic bin radius is fixed at $2.0<\tau<2.25$.
	}
\end{figure*}

The dependence of the signal upon the assumed ring morphology is further illustrated in Figure \ref{fig:LAT_Scans}. Here, we show the significance of the LAT excess counts as a function of $\tau$, $\zeta$, and $\phi$, by keeping $\phi$ or $\tau$ fixed and scanning the other parameters, and using a polar plot to demonstrate the $\phi$-dependence. This figure too indicates that the signal is particularly strong for the nominal ring parameters inferred from VERITAS and for $2.0<\tau<2.25$.

In conclusion, we find a significant, $\sim 3.4\sigma$ LAT excess at the same location and morphology as indicated by the VERITAS signal.
The values of the three ring morphology parameters ($\zeta\simeq 2.5$, $\phi\simeq 0\dgr$, and $\tau\simeq 2.1$) that maximize the significance of the LAT signal are similar to those that maximize the significance of the VERITAS signal, although in the latter only a lower limit on $\zeta$ could be established.

\subsection{Signal modelling}

To model the signal and better quantify its significance, we use a maximal likelihood (minimal $\chi^2$) analysis.
First, for given $\epsilon$ band and $\tau$ bin, we compute the $\chi^2$ contribution of the excess counts $\Delta n(\epsilon,\tau)$ with respect to the model prediction $\model(\epsilon,\tau)$,
\begin{equation} \label{eq:ChiSquared}
\chi^2(\epsilon,\tau,\Mbin)=\frac{\left(\Delta n - \model\right)^2}{f + \model} \fin
\end{equation}
The likelihood $\mathcal{L}$ is then related to the sum over all spatial bins and energy bands, as
\begin{equation}
\label{eq:Likelihood}\ln\mathcal{L} = -\frac{1}{2}\sum_{\epsilon,\tau}\chi^2(\epsilon,\tau) \fin
\end{equation}
The test statistics \citep{MattoxEtAl96_TS} TS, defined as
\begin{equation}
\myTS \equiv -2\ln\frac{\mathcal{L}_{max,-}}{\mathcal{L}_{max,+}}=\chi^2_{-} - \chi^2_{+} \coma
\end{equation}
can now be computed.
Here, subscript $-$ (subscript $+$) refers to the likelihood without (with) the modelled signal, maximized over any free parameters.

We examine both the shell model and the planar model for CRE injection, as discussed in \S\ref{sec:Theory}.
For \gama-rays, where the extent of the emission layer is negligible with respect to the bin size, each model has two free parameters, determining the location of the shock and the efficiency of CRE injection.
We choose these parameters as the semiminor axis of the projected shock surface, $b_s=\varrho_{b,s}$, and the logarithmic brightness $\epsilon J_\epsilon$ in the $2.0<\tau<2.25$ elliptical bin.

It is useful to parameterize the injection efficiency also in terms of the logarithmic flux $\epsilon F_\epsilon$ from the entire virial shock.
Using estimates based on the $\beta$-model, found in equations (A10) and (A14) of {\Stack}, this flux is given by
\begin{eqnarray} \label{eq:ICFluxBeta}
\epsilon F_\epsilon
& \simeq & \frac{1.0\times 10^{-13}}{(1+z)^{4}} \left(\frac{\xi_e\dot{m}}{0.01}\right) \left(\frac{\theta_{500}}{0\dgr.2}\right)^2  \left(\frac{r_s}{2R_{500}}\right)\left[\frac{H(z)}{H_0}\right]^{\frac{7}{3}} \nonumber\\
& & \times \left(\frac{M_{500}}{10^{14}M_\odot}\right)^{\frac{1}{3}}  \left(\frac{k_B T}{5\keV}\right) \erg \se^{-1} \cm^{-2} \fin
\end{eqnarray}
This result holds for both shell and planar models, providing an estimate of $\xi_e \dot{m}$.

For both models, the likelihood is estimated in the $1.5=\tau_{min}<\tau<\tau_{max}=8$ range, to avoid spurious contamination from the poorly modeled central region and from interfering structure at large radii.
We use all three energy bands, implicitly assuming a flat, $p=2$ injected CRE spectrum.
Changing $p$ to slightly softer, $\sim 2.2$ values, or incorporating also the fourth, photon-deprived high-energy band, introduces only a mild change in the resulting model parameters.

The best fit profiles of both models are shown in Figure \ref{fig:LAT_significance}.
In the first, shell model, we obtain $\myTS\simeq4.9$, corresponding to a $1.7\sigma$ signal. The best fit parameters here are $b_s=(2.3\pm0.1) \theta_{500}\simeq 1\dgr.5\pm0\dgr.1$, and $\epsilon J_\epsilon=3.3^{+2.7}_{-1.7}\times10^{-10} \erg \se^{-1} \cm^{-2} \sr^{-1}$, or equivalently $\epsilon F_\epsilon = 6.9^{+3.2}_{-3.0}\times10^{-9} \erg \se^{-1} \cm^{-2}$. In terms of CRE injection rate, this may be written (using Eq.~\ref{eq:ICFluxBeta}) as $\xi_e \mdot = 0.31^{+0.15}_{-0.14}$.

The second, planar model presents with $\myTS=8.9$, corresponding to a $2.5\sigma$ detection.
This is higher than in the shell model, because the planar model better captures the narrow (in $\tau$) signature.
The best fit parameters here are $b_s=2.14^{+0.07}_{-0.06}\times \theta_{500}\simeq 1\dgr.45^{+0\dgr.05}_{-0\dgr.04}$, and $\epsilon J_\epsilon=(5.5\pm1.8)\times10^{-10} \erg \se^{-1} \cm^{-2} \sr^{-1}$, or equivalently $\epsilon F_\epsilon = (4.0\pm1.3)\times10^{-9} \erg \se^{-1} \cm^{-2}$.
The acceleration efficiency can be determined from Eq.~(\ref{eq:ICFluxBeta}) in this model too; this yields $\xi_e \mdot = 0.19\pm0.07$.

\subsection{Robustness and comparison with previous studies}
\label{sec:LATRobustness}

Convergence and sensitivity tests for the foreground estimation and binning procedures were demonstrated in {\Stack}.
Overall, here we find that our results are robust even under significant changes to the polynomial fit order $N_f\ge0$ and to the angular extent $\tau_{max}>6$ of foreground estimation, are converged for HEALPix order $N_{hp}>9$ and under splitting the energy range to various $N_\epsilon>1$ bands, and are well-behaved for modest variations in $\Delta \tau$, $\tau_{min}$, and $\tau_{max}$.

\citet{FermiComa16} reported residual (background subtracted) emission from an area that overlaps partly with the Coma virial radius.
They placed $95\%$ one-sided (henceforth) upper limits in the range $(3.2\till5.8)\times 10^{-9}\se^{-1}\cm^{-2}$, depending on the template and spectrum assumed, on the flux $F(>100\MeV)$ of extended emission.
\citet{ZandanelAndo14} placed upper limits in the range  $(2.5\till2.9)\times10^{-9} \se^{-1}\cm^{-2}$ on $F(>100\MeV)$ using virial ring templates (ring, disc, and east--west ellipse).
Using a thick ($0\dgr.5$) elliptic ($\zeta=1\till3$) ring template, \citet{Prokhorov14} placed upper limits in the range $(2.4\till4.0)\times10^{-9} \se^{-1}\cm^{-2}$ on $F(>100\MeV)$.

In order to compare our results with these studies, we compute the $>100\MeV$ photon flux integrated over the entire virial shock.
In the shell model, we obtain $F(>100\MeV)=(6.9\pm3.1)\times10^{-9} \se^{-1}\cm^{-2}$, whereas in the planar model, which provides a better fit to the data, $F(>100\MeV)=(4.0\pm1.3)\times10^{-9} \se^{-1}\cm^{-2}$.
Our results are therefore comparable, and in some variants quite consistent, with these previous upper limits.
Moreover, such upper limits are sensitive to the precise template assumed, as demonstrated by the scatter among these studies and within each study; they are also quite sensitive to the modeled foreground \citep{FermiComa16}.
In particular, none of the previous studies modelled a thin ring, as we identify in LAT data.

To demonstrate this, we repeat the thick ring analysis of \citet{Prokhorov14}, centered on $\tau=1\dgr.3$ with width $\Delta \tau=0\dgr.5$. We do not identify a signal here, despite having a few more years of data and using our robust foreground removal method.
This imposes a one-sided $95\%$ upper limit of $F(>100\MeV)<2.0\times 10^{-9}\se^{-1}\cm^{-2}$, consistent, and even somewhat stronger, than the $F(>100\MeV)<3.3\times 10^{-9}\se^{-1}\cm^{-2}$ upper limit imposed by \citet{Prokhorov14} using the same template.
Hence, allowing for narrow structures in our analysis, as anticipated in \S\ref{sec:Theory}, is instrumental in detecting the signal.
This allows the flux of the narrow ring to exceed the flux upper limit based on a thick ring, by a factor of $\sim2$ for the planar model.

Foreground removal can have an important effect on our results.
The signal we report constitutes about one third of the total coincident flux (see Figure \ref{fig:LAT_flux}).
Our analysis removes all but the sharpest features (with the designated elliptical morphology), attributing weak gradients to the foreground.
It is likely that our analysis --- as well as the previous upper limits mentioned above --- attribute part of the signal, in particular its smoother parts, to the foreground, which is primarily Galactic and not precisely known \citep[\eg][]{KeshetEtAl04_EGRB}.

This effect is obfuscated by template misalignment and model uncertainties.
Our purely elliptic ring template cannot be assumed to precisely capture the full pattern of the shock.
As the template is thin, a small misalignment with respect to the projected shock would suffice for a substantial loss of flux, being attributed to the foreground rather than to the signal.
Due to this effect, the above flux estimates may somewhat underestimate the actual signal.

\section{ROSAT analysis}
\label{sec:ROSAT}

\begin{figure*}
	\centerline{
    \epsfxsize=8.5cm \epsfbox{\myfig{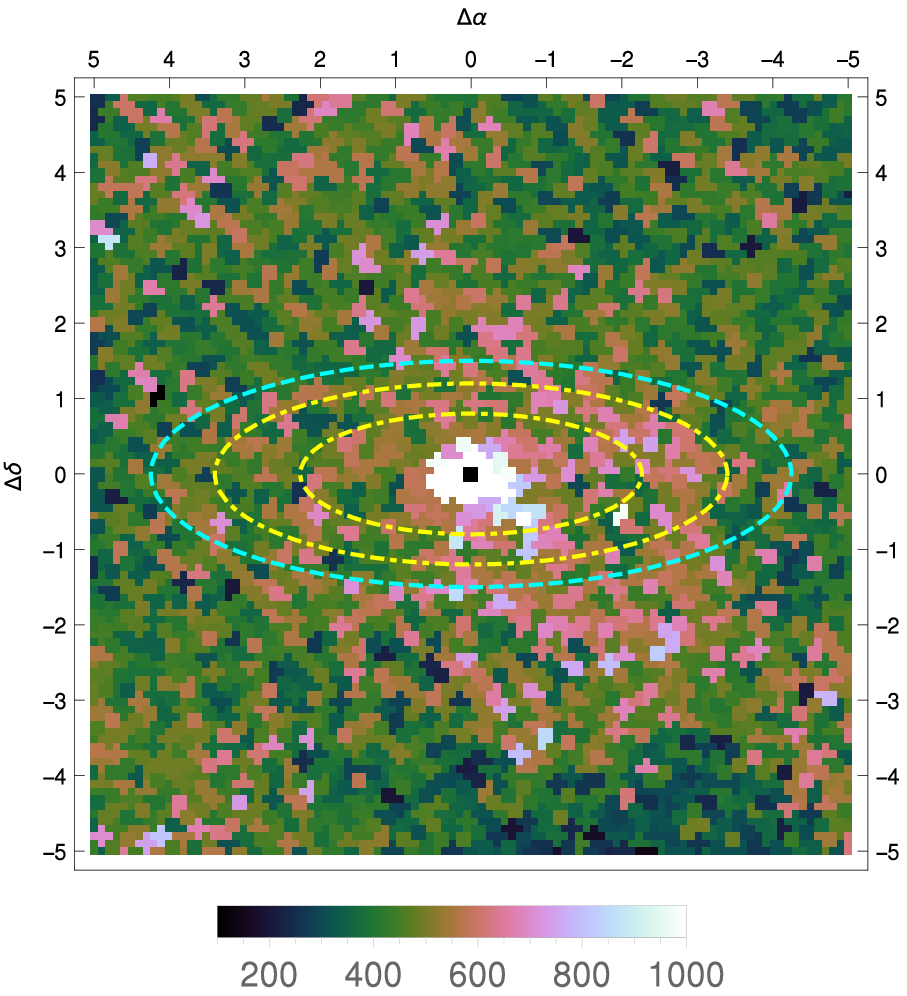}}
    \epsfxsize=8.5cm \epsfbox{\myfig{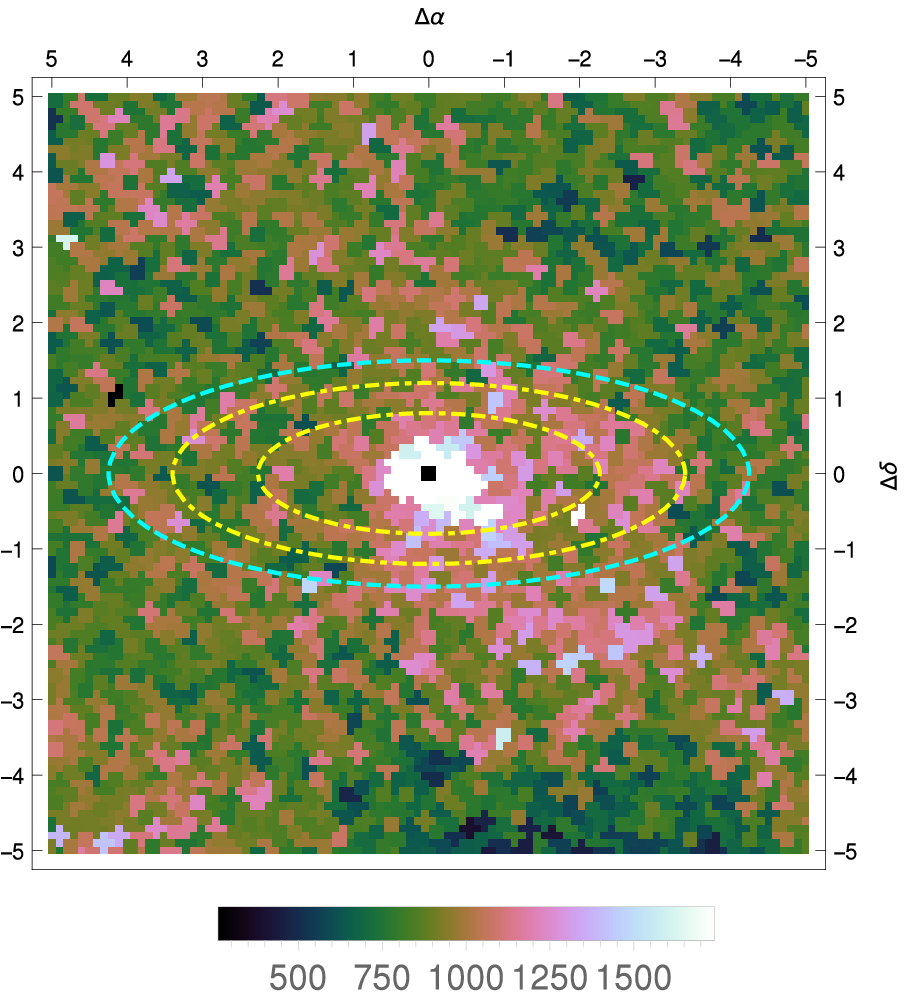}}
}
\caption{\label{fig:ROSATMaps}
Brightness in ROSAT bands R1 (left) and R1+R2 (right), in $10^{-6}\se^{-1}\mbox{ arcmin}^{-2}$ units.
The maps are centered upon Coma, in a CAR projection and equatorial coordinates.
Also shown are the LAT-based, $b_s\simeq 1\dgr.5$ shock position (dashed cyan contours) and the $\varrho_b\simeq 0\dgr.8$ and $1\dgr.2$ contours (dot-dashed yellow, as a guide to the eye) that enclose most of the X-ray virial signal, for our nominal, east--west ($\phi=0\dgr$), elongated ($\zeta=2.5$) ring morphology.
Here too, the signal is not sufficiently brighter than the foreground to be easily discernable by eye.
}
\end{figure*}

\subsection{Data preparation and analysis}
\label{sec:Data}

We use the \emph{ROSAT} all sky survey \citep[RASS;][]{SnowdenEtAl97}, with the Position Sensitive Proportional Counter (PSPC) of the X-ray telescope (XRT).
The provided\footnote{http://hea-www.harvard.edu/rosat/rsdc.html} PSPC maps were binned onto $12'\times12'$ pixels, well above the native $1'.8$ radius for $50\%$ energy containment.
Point sources were removed to a uniform source flux threshold for which their catalog is complete over $90\%$ of the sky \citep{SnowdenEtAl97}.
The exposure time in the Coma region is $\sim500\se$.

The $j_\epsilon\propto \epsilon^0$ spectrum of the thermal bremsstrahlung X-ray emission is harder than the $j_\epsilon\propto \epsilon^{-p/2}\sim \epsilon^{-1}$ spectrum of the virial shock signal, so it is advantageous to search for the latter at low energies.
In addition, for a steep density profile at the cluster outskirts, where adiabatic compression is substantial, the anticipated, binned virial signal is more pronounced at lower energies (see Figure \ref{fig:VSsingature}).
Therefore, to pick up the virial shock signal while minimizing the thermal contamination from the cluster and nearby structure, we focus on the lowest energy band, R1, spanning the energy range $0.110\till0.284\keV$. To test the signal, we also examine the next, R2 band, spanning the range $0.140\till0.284\keV$.

Maps centered upon the X-ray peak in Coma were retrieved from SkyView \citep{McGlynnEtAl98}.
The maps span $20\dgr$ with a $6'$ resolution in declination, with a rectangular (CAR) projection, so the solid angle per pixel is $\Omega_0\simeq2.7\times 10^{-6}\sr$.
This ensures a proper sampling of the binned map without approaching the instrumental resolution.
Next, we bin the X-ray counts in elliptic concentric annuli around Coma, following the preliminary signal from VERITAS and the aforementioned results from the LAT.
Zoomed in, $10\dgr$ maps of the brightness in the R1 and R1+R2 bands are shown in Figure \ref{fig:ROSATMaps}, along with contours  illustrating the elliptic binning.
Here, we define the R1+R2 band as the co-addition of bands R1 and R2 after normalizing each band to the same mean flux level, to allow equally weighted contributions from each band.

\subsection{Results}

\begin{figure}[b]
	\centerline{\epsfxsize=8.5cm \epsfbox{\myfig{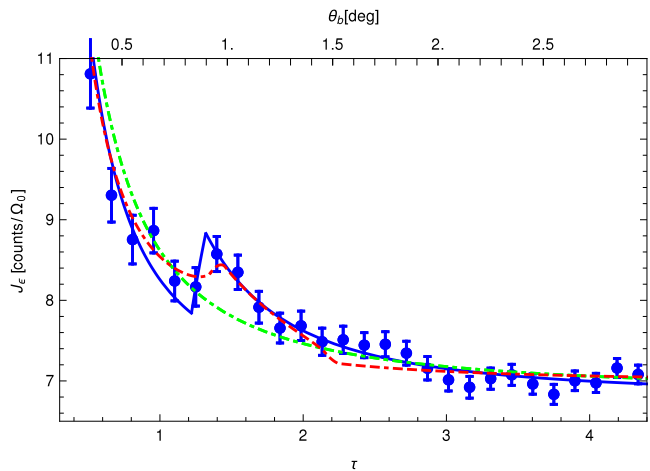}}}
\caption{\label{fig:ROSAT1Phi0}
Binned brightness profile around Coma in the ROSAT R1 band (blue error bars), for the nominal ring morphology.
Also shown are best fit models without (dot-dashed green) a virial CRE component, and with a virial signal in the shell model (dashed red), and in the planar model (solid blue).
}
\end{figure}

Figure \ref{fig:ROSAT1Phi0} shows the radial profile of the brightness in the R1 band for the nominal, east--west ($\phi=0\dgr$) elongated ($\zeta=2.5$) binning.
The central region of the cluster is dominated by diffuse thermal emission.
One can crudely model this component by fitting the binned profile as the combination of a uniform foreground and a thermal component with a radial power-law,  $c_0+c_1\varrho^{-a}$.
The best fit (shown as a dot-dashed green curve) indicates a projected power-law index $a\simeq 1.5$ (in band R1; $a\simeq 1.6$ in the combined bands R1+R2), which corresponds to a non-projected $j_\eps\propto r^{-2.5}$ emissivity profile.
This is somewhat flatter than anticipated by extrapolating the $\beta$-model to large radii, in part due to the known substructure $\sim 0\dgr.5$ southwest of the cluster's center, and in part due to the virial ring signal, as we show below.

\begin{figure*}
	\centerline{
        \epsfxsize=8.5cm \epsfbox{\myfig{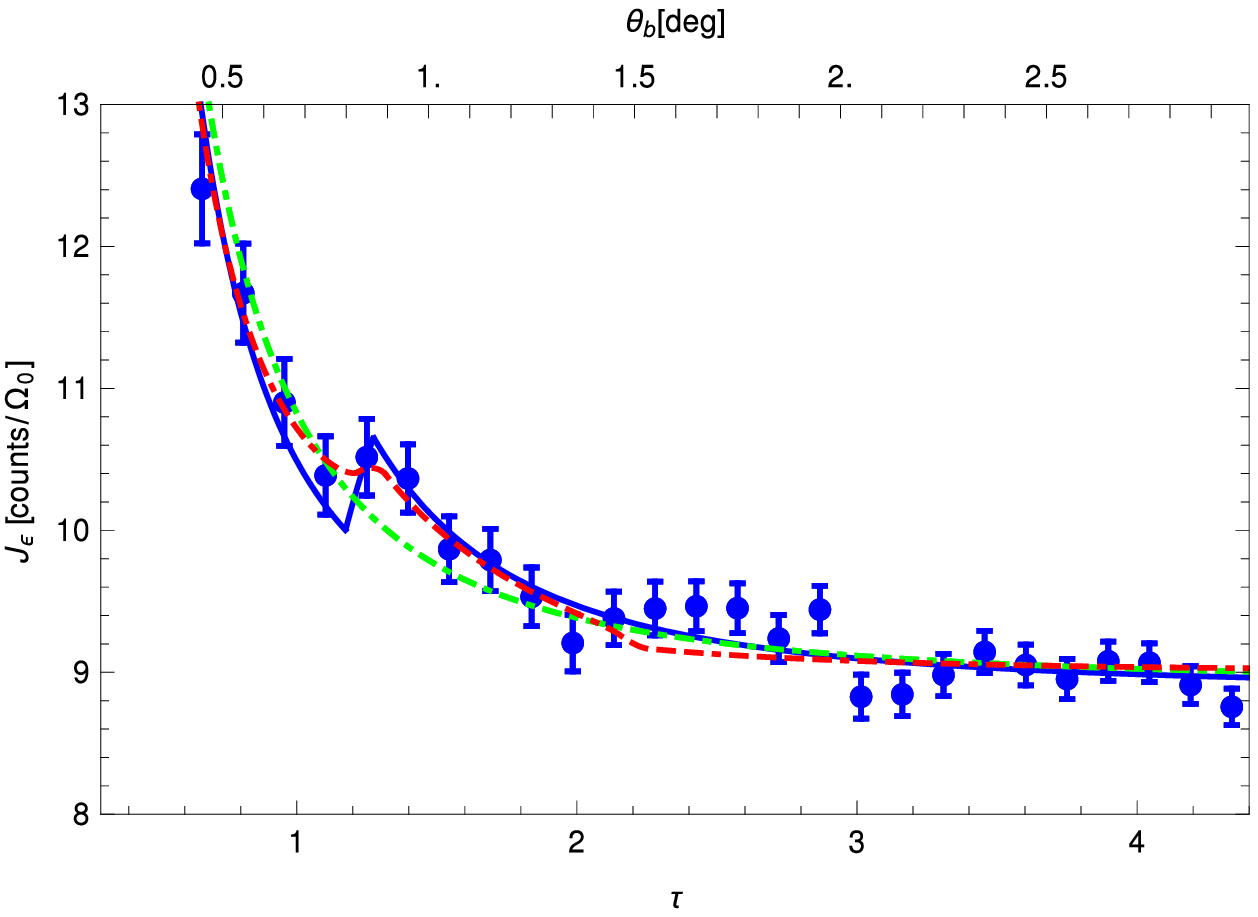}}
    %}
    %\centerline{
        \epsfxsize=8.5cm \epsfbox{\myfig{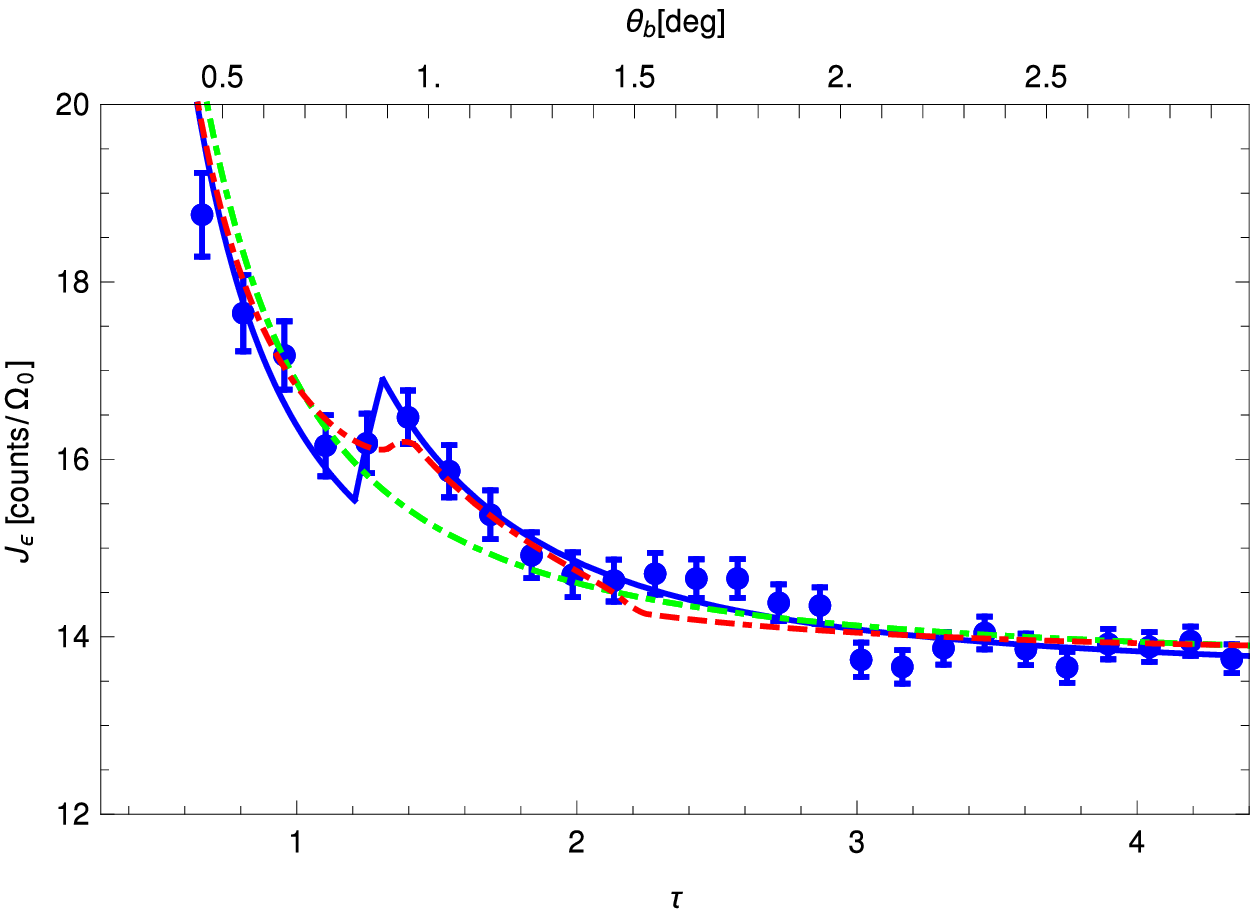}}
    }
\caption{\label{fig:ROSAT1and12Phi0}
Brightness profiles of ROSAT bands R2 (left) and R1+R2 (right). Notations and binning parameters are the same as in Figure \ref{fig:ROSAT1Phi0}.
}
\end{figure*}

At $\tau\simeq 1.5$, corresponding to a projected semiminor axis length $\varrho_b\simeq 1.5R_{500}$, or equivalently a semiminor separation angle $\theta_b\simeq 1\dgr$, some additional diffuse excess emission can be seen, in particular in the western side of the cluster.
This excess can be modelled as the emission from virial shock-accelerated CREs according to the binned analysis of \S\ref{sec:Theory}.
We consider both the shell model (Eqs.~\ref{eq:JModelBinned}--\ref{eq:VSB2}) and the planar model (Eqs.~\ref{eq:JModelPlanar}--\ref{eq:VSBPlanar}).
We first adopt the (normalized, semiminor) shock radius $\tau_s=2.2$, as inferred from the LAT analysis of \S\ref{sec:LAT}, and the nominal $q=8/3$ compression index corresponding to the isothermal sphere or extrapolated $\beta$-model gas distributions.
This leaves two free parameters, pertaining to the injection of CREs in the shock and to their evolution downstream.
The former constitutes an overall normalization factor, which can be chosen as the parameter $A$ of Eq.~(\ref{eq:ADef}) or $\tilde{A}$ of Eq.~(\ref{eq:APlanarDef}).
The latter parameter can be chosen as $\varrho_{cool}$, the (projected, semiminor) radius where the virial shock-accelerated CREs emitting in the ROSAT band typically  cool.

We thus fit the data with a two-parameter model, considering the shell model and the planar model separately, and quantify the likelihood of each model using the two-parameter confidence level corresponding to its TS statistics.
The resulting fit for the shell model, shown in Figure \ref{fig:ROSAT1Phi0} as a dashed red curve, presents at the $4.6\sigma$ confidence level, with  $\varrho_{cool}=1.4\pm0.2$.
Fitting the planar model, instead, gives a higher significance, $5.4\sigma$ signal (blue solid curve), with $\varrho_{cool}=1.27\pm0.05$.
These results are not sensitive to small changes in $\tau_s$, and vary smoothly with changes in $q$.
When allowing for the virial emission, the estimated slope of the inner profile steepens slightly, to $a\simeq 1.7\till1.9$.

The same analysis is next applied to the R2 band, and to the combined, R1+R2 band, as shown in Figure \ref{fig:ROSAT1and12Phi0}.
Band R2 shows essentially the same features as the R1 band, but as expected, the virial signal is weaker; it presents at a $2.5\sigma$ ($3.3\sigma)$ confidence level for the shell (planar) model, and the uncertainty in the model parameters is substantial.
The combined, R1+R2 analysis shows a significant signal for the shell model, at the $4.5\sigma$  confidence level, with $\varrho_{cool}=1.4\pm0.2$.
The planar model shows a higher significance, $5.7\sigma$ signal, with $\varrho_{cool}=1.26\pm0.05$.
The higher energy ROSAT bands, R4 ($0.44\till1.01\keV$) and R5--R7 ($>0.56\keV$), do not show a similar signal.

\subsection{Ring morphology}

\begin{figure}[h!]
	\centerline{
\epsfxsize=8.5cm \epsfbox{\myfig{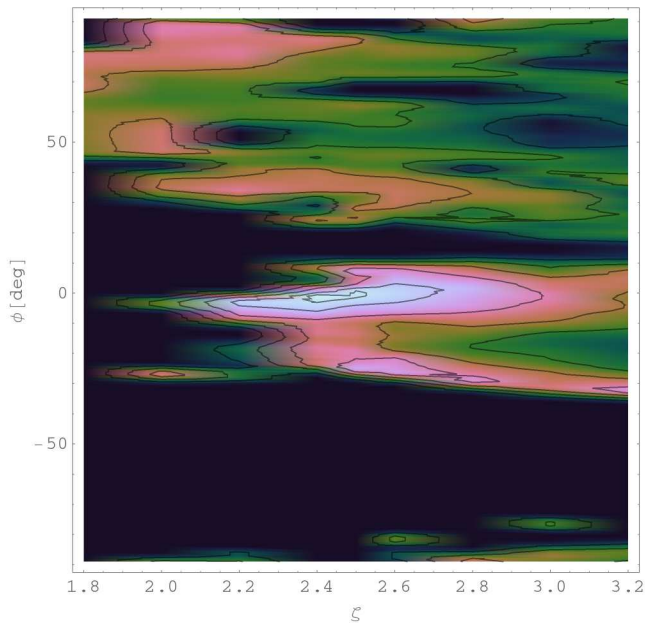}}
}
\caption{\label{fig:ROSATScan}
The TS-based significance of the planar model in the ROSAT band R1+R2. Contours are plotted at the $\{1, 2, 3, 4,5,5.5\}\sigma$ confidence levels.
We use a $\sigma_\phi=1\dgr$ smoothing and a $\chi_-^2/\nu<2.5$ threshold.
Here we assume $\tau_s=2.2$, but the results change very little in the relevant, $1.5<\tau_s<2.5$ range.
}
\end{figure}

The ring morphology parameters $\zeta$ and $\phi$ can be determined based on the X-ray signature alone.
Consider the planar model, which provides a better fit to the ROSAT data (as well as to the LAT data).
Figure \ref{fig:ROSATScan} shows the TS-based significance of the virial signal in the R1+R2 band, as a function of $\zeta$ and $\phi$.
The signal is noticeably localized near the nominal ring morphology.
The maximal significance is $6.3\sigma$, with  $\zeta=2.35_{-0.26}^{+0.30}$ and $\phi=0\dgr.7_{-7\dgr.4}^{+1\dgr.3}$ encompassing the nominal parameters.

Here we assumed that the shock lies at $\tau_s=2.2$, but the results remain nearly unchanged throughout the relevant,  $1.5<\tau_s<2.5$ range. This renders the evaluation of the ring morphology robust, but precludes a determination of the shock position based on the X-ray data alone unless the flow is accurately modelled.
A similar maximum near the nominal ring parameters is also found using band R1 alone, but this maximum is only local, with small other maxima emerging, including a narrow peak around $\phi\simeq -35\dgr$.

In Figure  \ref{fig:ROSATScan} we imposed a $\chi_-^2/\nu<2.5$ threshold, where $\nu$ is the number of degrees of freedom, to avoid contamination by poor $\mathcal{L}_-$ likelihood models for the central signal.
Lowering this threshold down to $1.5$ leaves a pronounced global maximum near the nominal ring parameters, but lowering it much further would leave no viable solution.
Raising the threshold much beyond 3 would introduce spurious local maxima.

\begin{figure}[h!]
	\centerline{
\epsfxsize=4cm \epsfbox{\myfig{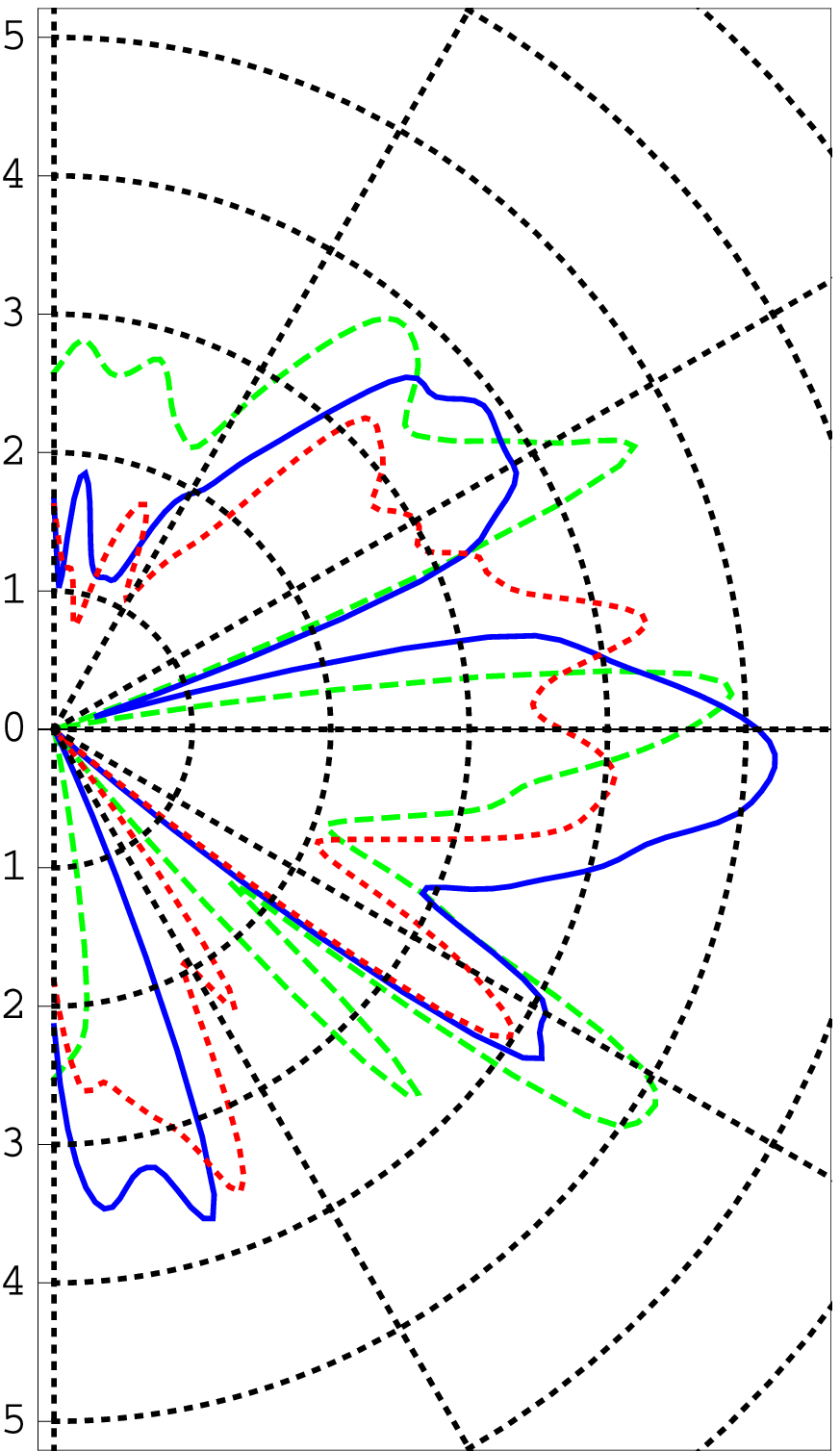}}
\epsfxsize=4cm \epsfbox{\myfig{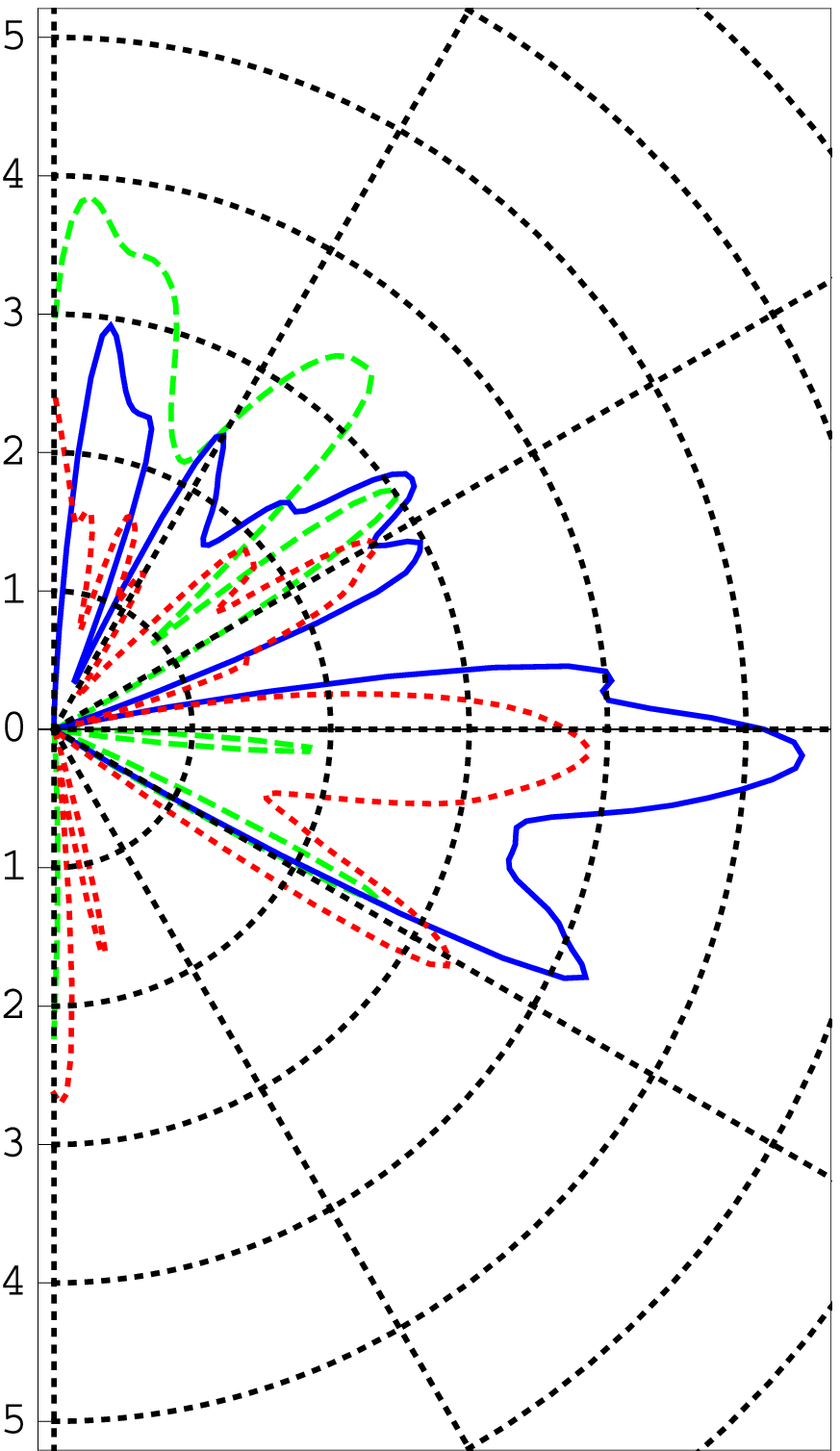}}}
\caption{\label{fig:ROSATPolar}
Polar plots of the TS-based significance (in units of standard deviation) in the planar model, as a function of $\phi$, in ROSAT bands R1 (left) and R1+R2 (right), for different elongations $\zeta$ (notations are the same as in the polar Figure \ref{fig:LAT_Scans}).
We use a $\sigma_\phi=2\dgr$ smoothing and a $\chi_-^2/\nu<2.5$ threshold.
}
\end{figure}

The ring morphology is also illustrated using polar plots, in Figure \ref{fig:ROSATPolar}.
Both R1 and R1+R2 bands show a strong signal maximized at the nominal ring parameters.
In band R1, this maximum is only local, and would not have been uniquely identified in the absence of the VERITAS prior.
However, it becomes a pronounced global maximum in the combined R1+R2 bands.

The emergence of the nominal ring parameters as a pronounced (and global, in the R1+R2 band) maximum of significance in X-rays alone, like it did in the VERITAS data and in the LAT data, indicates that the three signals reflect the same phenomenon.
The good agreement of the data with the model supports the interpretation of the signal as arising from virial shock-accelerated CREs, perceptibly advected downstream in the X-ray emitting regime.
The better agreement of the ROSAT data with the planar model than with the shell model, as found also for the LAT data in \S\ref{sec:LAT}, provides additional evidence for preferential accretion in the plane of the sky.

\section{Broadband analysis}
\label{sec:Combined}

\begin{figure*}
	\centerline{
        \epsfxsize=8.5cm \epsfbox{\myfig{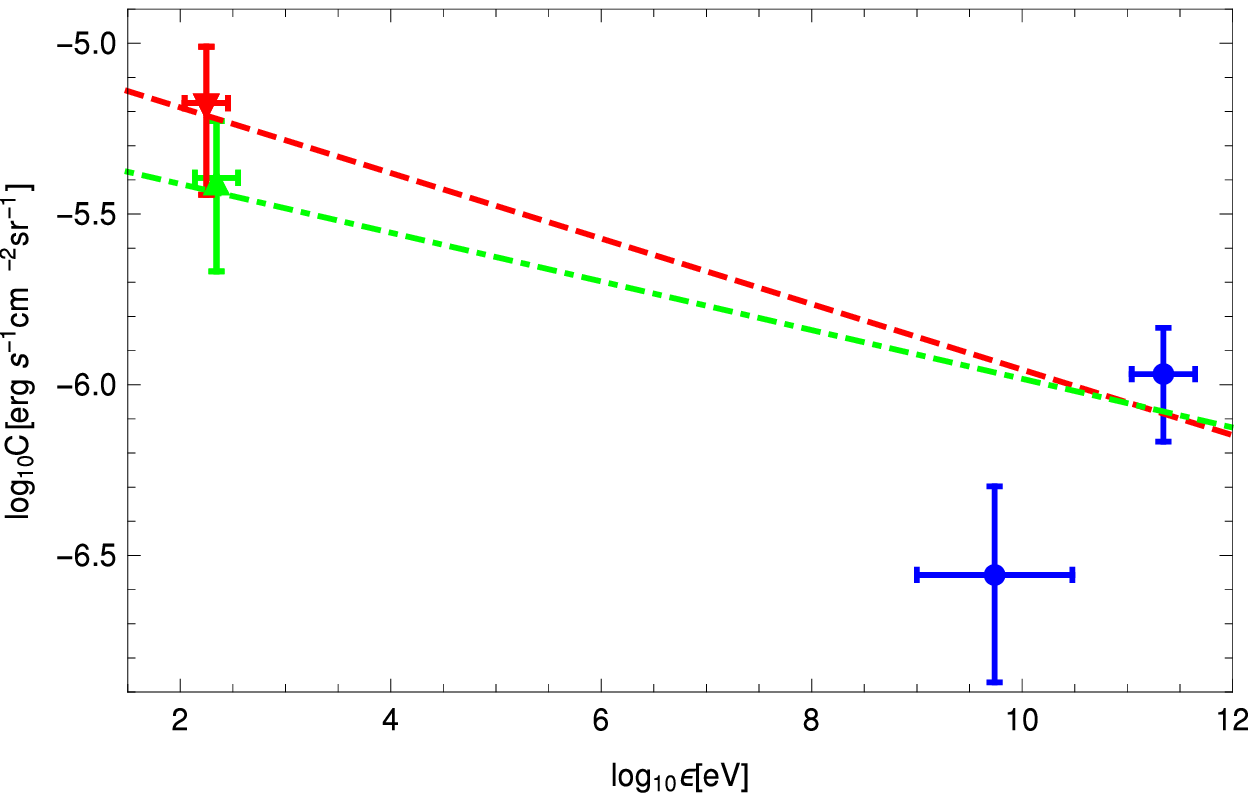}}
        \epsfxsize=8.5cm \epsfbox{\myfig{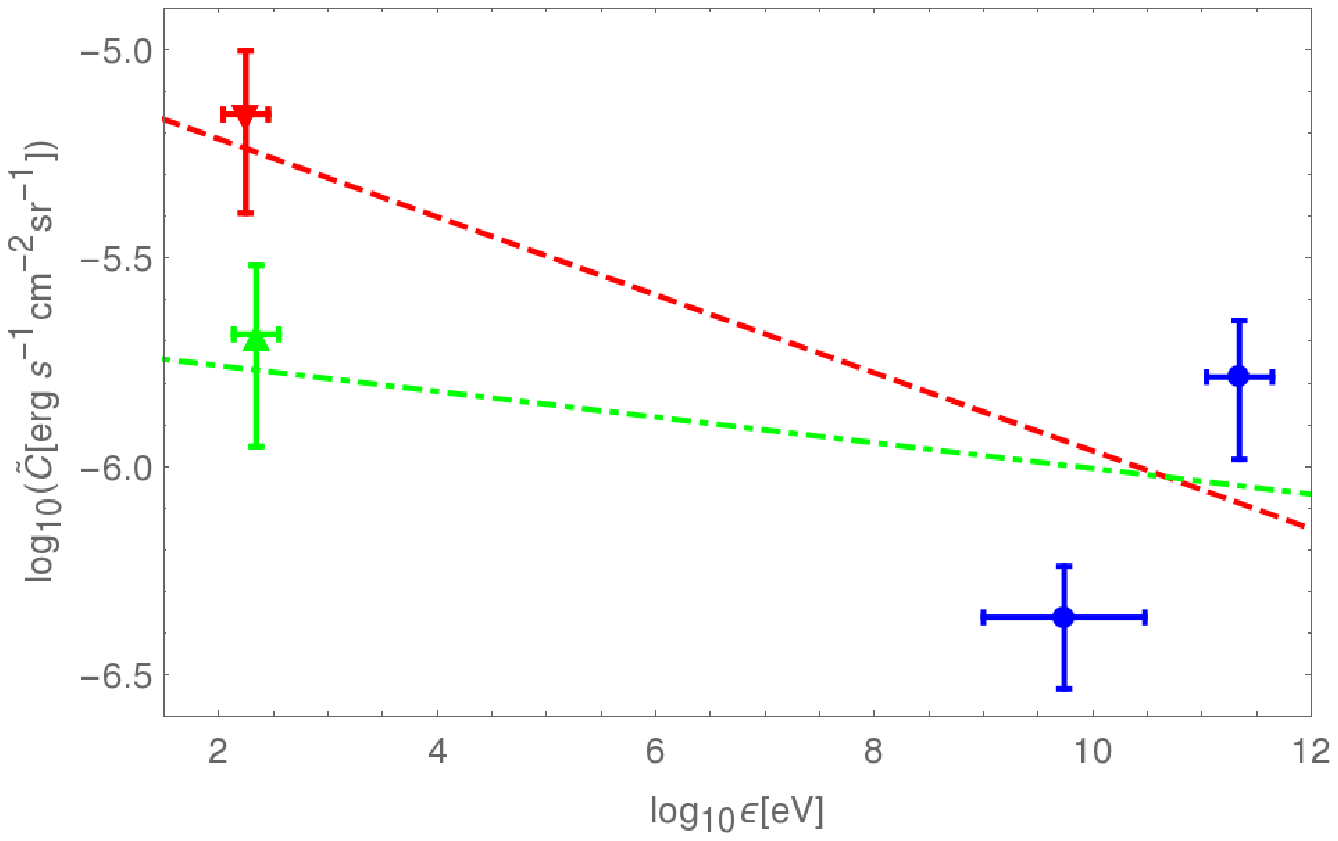}}
    }
    \caption{\label{fig:Broad_sig}%
    Injected CRE spectrum inferred from a broadband analysis in the shell (left panel) and planar (right panel) models.
    Shown are the injection normalizations $C=\eps_{\mbox{\tiny{GeV}}}^{-1/2}A$ (shell model) or $\tilde{C}=\eps_{\mbox{\tiny{GeV}}}^{-1/2}\tilde{A}$ (planar model), derived separately in the three bands (ROSAT R1, LAT $1\till30\GeV$, and VERITAS $\sim220\GeV$; error bars), along with the best fit power-law fits (curves).
    The values of $C$ and $\tilde{C}$ inferred from ROSAT depend on the adiabatic compression index $q$; results are shown for an isothermal sphere ($q=8/3$; red down triangle; dashed curve) and for a steep gradient ($q=16/3$; green up triangle, slightly offset in $\eps$ for visibility; dot-dashed). The energy range plotted for VERITAS is only representative, as the effective area has a non-trivial energy dependence and the upper limit is not well constrained.
}
\end{figure*}

We may now combine the virial signals outlined above and in {\Coma} in order to measure the spectrum of CREs injected at the shock.
This spectrum is assumed to be a power-law, $dN_e/dE\propto E^{-p}$.
The CREs emitting \gama-rays cool rapidly, before they can travel far from the shock, so the VERITAS and LAT signals directly trace the cooled CRE spectrum, $J_\eps=\eps\, dn_\gamma/d\eps\propto \eps^{-p/2}$.
However, these two \gama-ray telescopes do not span a sufficiently wide range of photon energies $\eps$ to permit a good spectral measurement, considering the substantial uncertainties in flux determination, and the different systematics of each telescope.
In contrast, the wide energy range spanned by a combination of \gama-ray and X-ray measurements is sufficient for a spectral measurement.
However, X-ray emitting CREs do travel far from the shock, so one needs to take into account their propagation and compression.

Therefore, we evaluate the spectrum by first extrapolating each signal to the corresponding emissivity at the shock, before the radiating CREs can propagate and evolve, using the analysis of \S\ref{sec:Theory}.
The shock emissivity then directly yields the flux of CREs injected at the shock.
We carry out the analysis separately for the shell model and for the planar model; recall that the latter fits better both the LAT data and the ROSAT data.

For the shell model, we quantify the emissivity at the shock using the parameter $C\equiv (\eps/\eps_0)^{-1/2}A\equiv\eps_{\mbox{\tiny{GeV}}}^{-1/2}A$, where the normalization $A$ of Eq.~(\ref{eq:ADef}) is inferred from the binned analysis of Eqs.~(\ref{eq:JModelBinned}) and (\ref{eq:VSB2}).
Here, $\eps_0$ is a reference photon energy, which we arbitrarily choose as $1\GeV$.
Similarly, for the planar model we use the parameter $\tilde{C}\equiv\eps_{\mbox{\tiny{GeV}}}^{-1/2}\tilde{A}$, where the latter term is defined in Eq.~(\ref{eq:APlanarDef}) and is derived from the binned analysis of Eq.~(\ref{eq:VSBPlanar}).
The two parameters scale in the models as a power law in photon energy, $\{C,\tilde{C}\}\propto \epsilon^{-\eta}$, where $\eta\equiv{(p-2)/2}$.
Their definition is convenient, because they are energy independent for a flat ($p=2$) CRE spectrum.
Fitting their values as inferred from the data, as an $\eps^{-\eta}$ power-law, yields an estimate of the CRE spectrum injected at the shock: $p=2+2\eta$.

Figure \ref{fig:Broad_sig} shows the CRE injection estimates based separately on ROSAT, on Fermi-LAT, and on VERITAS.
The left panel depicts the values of $C$ inferred from the shell model, whereas the right panel shows the values of $\tilde{C}$ derived from the planar model.
To obtain more reliable estimates, we use the locally measured brightness $\eps J_\eps$, rather than the integrated flux $\eps F_\eps$.

In the VERITAS mosaic \citep{VERITAS12_Coma}, a thick ring spanning the semiminor axis range $1\dgr.0<\theta_b<1\dgr.6$ shows an excess signal of $410\pm150$ counts (\Coma).
Taking into account the observation duration $t\simeq 18.6\mbox{ hr}$ and the estimated effective area along with its energy dependence \citep{MaierEtAl08}, and assuming a flat, $p=2$ spectrum, this corresponds to a flux $\eps J_\eps=(7.5\pm 2.7)\times 10^{-10}\erg \se^{-1}\cm^{-2}\sr^{-1}$.
Notice that this flux is about half that of the rough estimate in {\Coma}, where a constant value was adopted for the typical effective area.
This corrected flux corresponds in the shell model, according to Eqs.~(\ref{eq:JModelBinned}) and (\ref{eq:VSB2}), to an injected CRE normalization $C=(1.1\pm 0.4)\times 10^{-6}\erg\se^{-1}\cm^{-2}\sr^{-1}$.
In the planar model, it corresponds to $\tilde{C}=(1.6\pm 0.6)\times 10^{-6}\erg \se^{-1}\cm^{-2}\sr^{-1}$.

For the LAT, the shell model indicates a flux $\eps J_\eps=3.3_{-1.7}^{+2.7}\times10^{-10}\erg\se^{-1}\cm^{-2}\sr^{-1}$ in the $2.0<\tau<2.25$ bin, in which the signal is significant. This corresponds to an injected $C=2.8_{-1.4}^{+2.3}\times10^{-7}\erg\se^{-1}\cm^{-2}\sr^{-1}$.
In the better fitting, planar model, the flux in this $\tau$ bin becomes $\eps J_\eps=(5.5\pm 1.8)\times 10^{-10}\erg\se^{-1}\cm^{-2}\sr^{-1}$, corresponding to $\tilde{C}=(4.3\pm 1.4)\times 10^{-7}\erg \se^{-1}\cm^{-2}\sr^{-1}$.

For ROSAT, the inferred values of $A$ and $\tilde{A}$ are sensitive to the level of adiabatic compression experienced by the CREs, and to the column density of absorbing gas, estimated as $9\times 10^{19}\cm^{-2}$ \citep{DickeyLockman90,KalberlaEtAl05}.
Here we use the R1 band.
For $q=8/3$, we obtain, using the unabsorbed flux according to the PIMMS \citep[v4.8d;][]{Mukai93} tool, $C=(6.7\pm3.1)\times   10^{-6}\erg \se^{-1}\cm^{-2}\sr^{-1}$ for the shell model, and $\tilde{C}=(7.0\pm3.0)\times 10^{-6}\erg \se^{-1}\cm^{-2}\sr^{-1}$ for the planar model.
For $q=16/3$, we find $C=(4.0\pm1.9)\times 10^{-6}\erg \se^{-1}\cm^{-2}\sr^{-1}$ and $\tilde{C}=(2.1\pm1.0)\times 10^{-6}\erg\se^{-1}\cm^{-2}\sr^{-1}$.

Fitting these X-ray through \gama-ray estimates as a single power-law, we find that in the shell model $p=2+2\eta=2-2d\ln C/d\ln \eps = 2.19\pm0.04$ for $q=8/3$ ($p = 2.14\pm0.04$ for $q=16/3$), shown as a dashed (dotted) curve in the left panel of Figure \ref{fig:Broad_sig}.
In the planar model, we find $p=2.19\pm0.04$ for $q=8/3$ ($p=2.06\pm0.04$ for $q=16/3$), shown in the right panel.
These results demonstrate how correcting for the rise in CRE energy density due to the adiabatic compression slightly hardens the reconstructed spectrum.
The boost in CRE energy $E$ due to adiabatic compression does not appreciably change the results.

The CRE spectrum injected at the shock, found to lie in the range $2\lesssim p\lesssim2.2$, is consistent with the nearly flat spectrum anticipated in a strong shock.
It is similarly consistent with the spectrum derived from the LAT stacking analysis (\Stack).
Conversely, this result supports the interpretation of the three signals as not only arising from the same mechanism, but also as inverse-Compton emission from CREs injected by the virial shock.

The CRE injection rate $\xi_e\dot{m}\simeq 0.2\%$, estimated above for the $\sim$few GeV LAT range by adopting the planar model and assuming $p=2$, changes only slightly, to $\xi_e\dot{m}\simeq 0.3\%$ (with an uncertainty factor of $\sim 2$), for $p\simeq 2.2$.
Here, we took into account CRE Lorentz factors in the range $1\lesssim \gamma\lesssim 10^8$ (the same result is obtained, \eg for $5\lesssim \gamma\lesssim 10^7$).

As Figure \ref{fig:Broad_sig} shows, the injection rate corresponding to the LAT signal is noticeably lower then that inferred from the VERITAS mosaic.
The two are inconsistent at the $1.8\sigma$ ($2.0\sigma$) level in the shell (planar) model, suggesting some systematic error in the analysis.
Some of the LAT signal may have been attributed to the stronger foreground, and some of it must have been missed due to the narrow template, as discussed in \S\ref{sec:LATRobustness}; correcting for such a putative effect would somewhat flatten the inferred spectrum, toward $p\simeq 2$.
Alternatively, the VERITAS analysis (\Coma) may be affected by systematic errors associated with the observational mode, which was not intended for observing a diffuse signal; a fainter VERITAS signal would imply a softer spectrum, $p\gtrsim 2.2$.
Further research is needed to resolve the discrepancy and facilitate a more accurate measurement of the spectrum.

\section{Summary and Discussion}
\label{sec:Discussion}

Following preliminary evidence (\Coma) for an elongated virial ring in a $\sim220\GeV$ VERITAS mosaic, we examine if there are coincident signals in $\gtrsim\GeV$ \gama-rays from Fermi-LAT (see Figure \ref{fig:ComaFermiMap}), and in soft, $\sim0.1\keV$ X-rays from ROSAT (Figure \ref{fig:ROSATMaps}).
Such emission is expected from the CREs accelerated by the virial shock, as they inverse-Compton scatter CMB photons.
The anticipated signature is approximately an ellipse, with a ratio $\zeta\equiv a/b\gtrsim2.5$ of semimajor axis to semiminor axis, a semiminor axis in the range  $1\dgr.0<b<1\dgr.6$, and an approximately
east-west, $\phi\simeq 0\dgr$ major axis orientation.

We analyze the broad-band signature of CREs injected by the virial shock (see Figure \ref{fig:VSsingature}).
The radius, thickness, and spectrum of the resulting, so-called virial ring emission, depend on the energy of the photon (or equivalently, of the emitting CRE).
In high, in particular LAT and VERITAS, energies, a maximally thin (bin- or PSF-limited) photon excess with a nearly flat spectrum ($\alpha\simeq 2$) is expected.
In low, in particular soft X-ray, energies, the ring should become smaller and thicker due to advection downstream.
Adiabatic compression amplifies this signal, leading to an apparent spectral softening.
The corresponding ROSAT signal is expected to surface above the foreground only at the lowest energy bands.

The broad-band signature is derived for non-binned (Eqs.~\ref{eq:JModel}--\ref{eq:VSB}) and binned (Eqs.~\ref{eq:JModelBinned}--\ref{eq:VSB2}) analyses, under the assumption of homogeneous CRE injection along the shock surface, in the so-called shell model.
However, the distribution of known LSS surrounding Coma (Figure \ref{fig:ComaEnv}) suggests that accretion through the virial shock may be particularly strong in the plane of the sky, motivating the introduction of a planar model (Eqs.~\ref{eq:APlanarDef}--\ref{eq:VSBPlanar}).
Here, the rings are thinner than in the shell model, and the signal vanishes at small radii.

In both LAT and ROSAT data, we find signals at the expected ring elongation $\zeta\simeq 2.5$, orientation $\phi\simeq 0\dgr$, approximate shock position $\tau_s$, and brightness $J_\eps$.
Each of these signals is not, on its own accord, and ignoring prior information, highly significant.
The LAT signal presents locally as a thin (PSF-broadened), $2.1\lesssim\tau\lesssim2.2$ elliptical ring, at the $3.4\sigma$ confidence level (Figures \ref{fig:LAT_flux} and \ref{fig:LAT_significance}).
However, fitting our model (with TS statistics, taking into account trial factors) indicates only a $2.5\sigma$ detection (for the planar model; $1.7\sigma$ for the shell model).
The ROSAT signal presents as a smaller-scale, $1.4\lesssim\tau\lesssim2$, extended excess (Figures \ref{fig:ROSAT1Phi0} and \ref{fig:ROSAT1and12Phi0}), at the $5.7\sigma$ confidence level (TS statistics in band R1+R2 for the planar model; $4.5\sigma$ for the shell model).
However, in band R1 alone, a comparably good fit can be found at a few other combinations of $\zeta$ and $\phi$ (Figure \ref{fig:ROSATPolar}).

Nevertheless, these signals, combined, indicate a virial shock signal at a high confidence level, because
(i) the VERITAS analysis has already pinpointed the ring parameters $\zeta$, $\phi$, and, with some uncertainty, also $\tau_s$ and $J_\eps$;
(ii) the combined significance of the three signals is very high, although they cannot be simply co-added due to the different analysis techniques and their systematics;
(iii) the LAT excess is maximized at approximately the ring parameters inferred from VERITAS (Figures \ref{fig:LAT_Scan} and \ref{fig:LAT_Scans});
(iv) the ROSAT significance too is maximized at the same parameters (Figures \ref{fig:ROSATScan} and \ref{fig:ROSATPolar}), globally (band R1+R2) or at least locally (R1);
(v) the brightness of the three signals agrees well with the expected flat, $p\simeq 2.0\till 2.2$ injected CRE spectrum (Figure \ref{fig:Broad_sig});
(vi) the inferred CRE acceleration efficiency qualitatively agrees with a previous independent estimate ({\Stack});
and
(vii) anecdotal evidence that the shock tracers  (ROSAT, LAT, VERITAS, synchrotron and SZ; see {\Coma}) are related, for example their  similarly dominant western part.

The broadband signal corresponds to a CRE injection rate $\xi_e\dot{m}\sim 0.3\%$ over a Hubble time.
Due to the unknown three-dimensional morphology, the differences between the models, the dependence upon additional processes such as CRE advection and diffusion, and the systematics inherent to each model and analysis pipeline, this can only be determined to within a factor of $\sim 3$.
This result compares favorably with the $\xi_e\dot{m}\sim 0.5\%$ (with a systematic uncertainty factor of $\sim 2$) estimate derived from a stacking analysis (\Stack) that combined LAT data around 112 other galaxy cluster.

We find that the planar model, in which CRE injection is assumed to preferentially take place in the plane of the sky, provides a better fit to both the LAT data and the ROSAT data, with respect to the shell model, in which injection is assumed to be uniform along a triaxial shock surface.
If substantiated, this would demonstrate how the leptonic virial shock signal can gauge the large scale environment of the cluster.
If such a planar signature is furthermore found to be typical of cluster virial shocks, then this may explain why the stacked LAT ring (\Stack) appeared to be slightly (but not significantly) narrower than expected.
Indeed, after the smooth gradients have been filtered out, there is a bias for preferentially picking out the sharper, face-on signature of planar emission.

Our LAT results are consistent with previous upper limits imposed on the \gama-ray emission from Coma \citep[\eg][]{ZandanelAndo14, Prokhorov14, FermiComa16}. Although our flux estimate slightly exceeds some of these limits, this can be traced to different template and background removal methods; see \S\ref{sec:LATRobustness}.
The elliptic ring-like signal appears considerably narrower in the LAT than it does in the VERITAS mosaic, where is presents with a (FWHM) thickness of $\sim 0\dgr.5$ (\Coma).
However, this is probably due to the observational mode used by VERITAS, in particular the $0\dgr.4$ integration diameter of the on-region. The comparison is further complicated by the wobble mode and ring background model used to produce the VERITAS mosaic, which are not intended for observing a diffuse signal.

CREs that radiate in the X-ray band have a considerable time to travel away from the shock, by advection and by diffusion; the latter could in principle smear and effectively remove a detectable signal.
The detection of a localized ROSAT signal thus places an upper limit on the CRE diffusion coefficient, $D$.
The best fit to the ROSAT signal indicates a $\varrho_{cool}= (1.26\pm0.05)R_{500}$ (semiminor) radius of cooling.
This corresponds to an upper limit $D(E\simeq 300\MeV)\lesssim10^{32}\cm^2\se^{-1}$, consistent with typical estimates for the ICM \citep[][and references therein; note that $D$ typically increases with energy]{Keshet10}.

The brightness of the X-ray signal depends on the level of adiabatic compression, which competes with diffusion and thus can provide some measure of $D$, but this would require better data than presently available.
The distribution of low-energy CREs should gradually approach the distribution of the X-ray gas, so one may expect less elongated and less planar emission in soft X-rays with respect to the \gama-ray signal. We cannot establish or rule out such an effect with the present data.

\acknowledgements
We thank G. Ilani, I. Gurwich, and D. Prokhorov for helpful discussions.
This research has received funding from the IAEC-UPBC joint research foundation (grant No. 257), and was supported by the Israel Science Foundation (grant No. 1769/15) and by the GIF (grant I-1362-303.7/2016).
We acknowledge the use of NASA's SkyView facility (http://skyview.gsfc.nasa.gov) located at NASA Goddard Space Flight Center.

\bibliography{Virial}

\end{document}